\documentclass[aps,prb,showpacs]{revtex4}
\usepackage{graphicx,amsmath,amssymb,color}
\newcommand{\mbb}{\mathbb}
\newcommand{\mc}{\mathcal}
\newcommand{\tet}{\texttt}
\newcommand{\ve}{\varepsilon}

\begin{document}
\title{Plasmon dissipation in gapped-graphene open systems at finite temperature}
\author{Andrii Iurov$^{1}$, Godfrey Gumbs$^{2,3}$, Danhong Huang$^4$, and V. M. Silkin$^3$}
\affiliation{$^{1}$Center for High Technology Materials, University of New Mexico, Albuquerque, NM 87106, USA\\
$^{2}$Department of Physics and Astronomy, Hunter College of the City University of New York, 695 Park Avenue, New York, NY 10065, USA\\
$^{3}$Donostia International Physics Center (DIPC), P de Manuel Lardizabal, 4, 20018, San Sebastian, Basque Country, Spain\\
$^{4}$Air Force Research Laboratory, Space Vehicles Directorate, Kirtland Air Force Base, NM 87117, USA}

\date{\today}

\begin{abstract}
Numerical and closed-form analytic expressions for plasmon dispersion relations and rates of
dissipation  are first obtained at finite-temperatures for  free-standing gapped graphene.
These closed-system results are  generalized to an open system with Coulomb coupling of
graphene electrons to an external electron reservoir. New plasmon modes, as well as new
plasmon dissipation channels, are found in this open system, including significant modifications
arising  from the combined effect of thermal excitation of electrons and an energy bandgap
in gapped graphene. Moreover, the characteristics of the new plasmon mode and the additional
plasmon dissipation may be fully controlled by  adjusting the  separation between the graphene
layer from the surface of a thick conductor. Numerical results for the thermal shift of plasmon
frequency in a doped gapped graphene layer, along with its sensitivity to the local environment,
 are demonstrated and analyzed.  Such phenomenon associated with the frequency shift of plasmons
 may be applied to  direct optical measurement of local electron temperature in transistors and
nanoplasmonic structures.
\end{abstract}

\pacs{73.21.-b, 71.70.Ej, 71.45 Gm, 73.20.Mf}
\maketitle

\section{Introduction}
\label{sec1}

Plasmon excitations in graphene are one of the most exciting and actively studied subjects
both theoretically\, \cite{Wun, DS07, pavlo, DSmain, Chakraborty, Gb, Rafael} and experimentally\,\cite{PoliNano, Poli1, Poli2, Poli3, Poli4}. Graphene plasmons are especially
important, partially because of their versatile frequency range tuned by varying doping
concentration and energy band gap. Consequently, graphene is expected to have several
potential device applications in transistors, optics, microscopy and
nanolithography\,\cite{RG1, RG2, RG3}. These studies on graphene plasmonics also extend
to other carbon-based structures, such as fullerenes\,\cite{Nord1, FO1, FO2, FO3, Anto, FOLiu},
carbon nanotubes\,\cite{nt1, nt2}, and especially  recently discovered silicon-based silicene
and other buckled honeycomb lattice structures\,\cite{silicene, SP}
with on-site potential differences between  sublattices.
\medskip

It is well known that the dynamics of self-sustaining density oscillations (i.e., plasmons)
in a closed system may  be very well described by Dyson's equation in many-body theory for
two-particle Green's functions, from which both the plasmon dispersion relation and the
intrinsic plasmon dissipation rate (inverse quasiparticle lifetime) may be determined \,\cite{Gb}.
For an open system\,\cite{dhh-2,dhh-3,dhh-4,dhh-5,dhh-6,dhh-7,dhh-8, Rec}, on the other hand, the dynamics
for carrier excitations is much more
complicated and will strongly depend on the electronic coupling to its environment (external
reservoir). Open-system dynamics  (both classical and quantum) includes tunnel-coupling of a
conductor to external electrodes \,\cite{dhh-9}, leakage-of an optical cavity to free space \,\cite{dhh-10},  and
thermal coupling of an electronic system to a heat bath \,\cite{dhh-11}. The coupling to an external reservoir
will introduce additional  dissipation channels for an electronic system, in addition to an extra
contribution for internal electron interactions. Such energy-dissipation dynamics in an open system
 has been  treated in the past using the so-called Lindblad dissipative
superoperator \,\cite{dhh-12}.

\medskip

\begin{figure}
\centering
\includegraphics[width=0.43\textwidth]{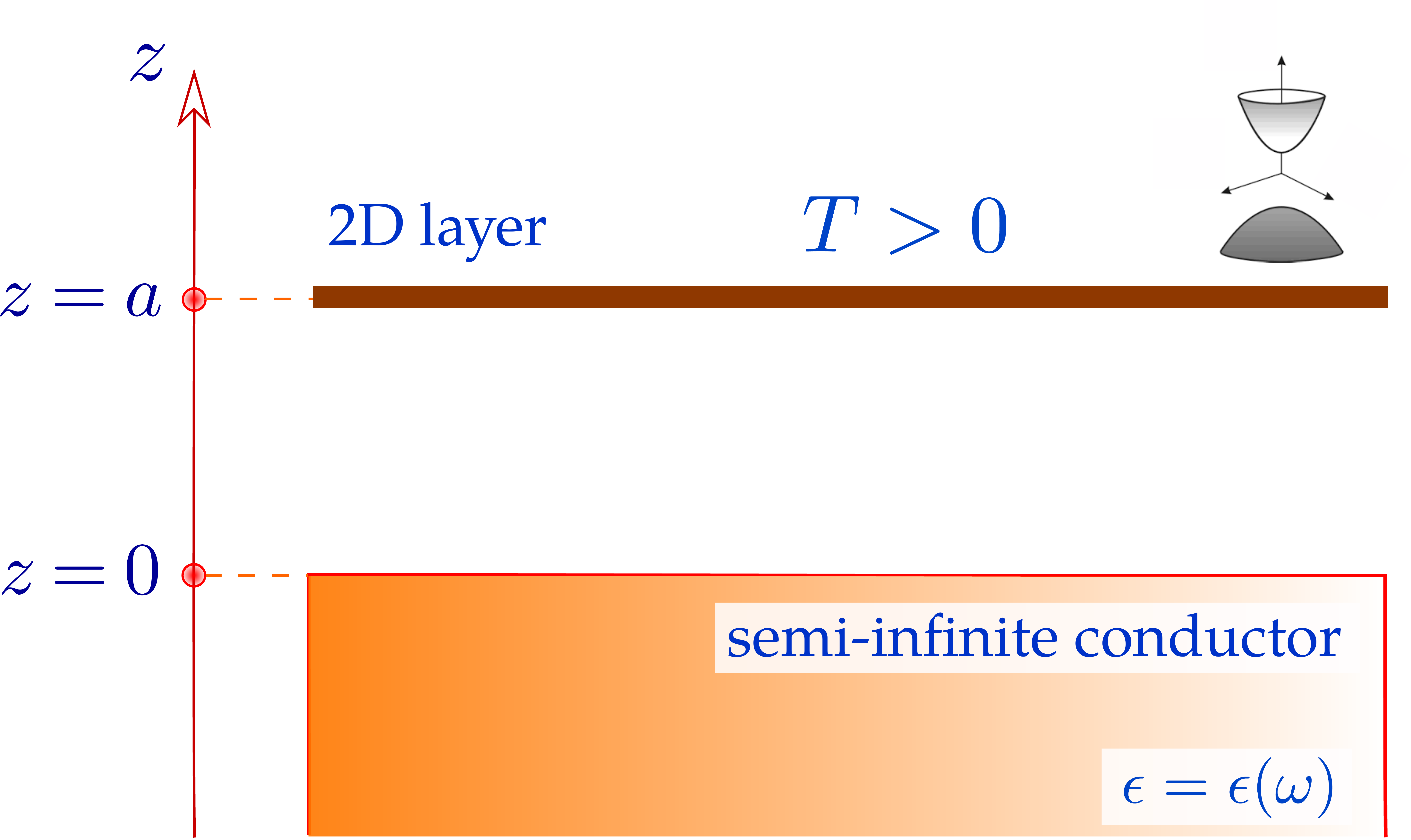}
\caption{(Color online) Schematics of a graphene-based hybrid structure (open system), which consists of
a gapped graphene layer coupled by Coulomb interaction to a semi-infinite conducting substrate (electron reservoir).
The graphene layer is separated by distance $a$ from the surface.}
\label{FIG:1}
\end{figure}

Although the plasmon dynamics in free-standing monolayer graphene- (closed system) has been
explored extensively, there are much fewer optical studies for graphene coupled by Coulomb
interaction to an external electron reservoir \,\cite{ours}. The existence of such an reservoir will introduce
Coulomb coupling between the graphene electrons and those carriers in a conducting substrate, as
illustrated in our proposed nanoscale hybrid structure in Fig.\,\ref{FIG:1}. Dielectric screening of the  electron-electron interaction within the graphene layer will be modified by their coupling
to the reservoir, thereby leading to a new plasmon mode as well as a new dissipation channel concurrently \,\cite{dhh-12}.
Moreover,  introduction of an energy gap in monolayer graphene as well as thermal excitations of
electrons at finite temperatures, will further modify this Coulomb coupling to the reservoir.
The main application of the current study aims to establish a basic principle for a contactless
measurement of local electron temperature. One way for non-invasive measurement of local electron
temperature \,\cite{dhh-14,dhh-15,dhh-16,dhh-17,dhh-18} is based on the thermal shift of plasmon 
energies, including the effects of graphene coupling to an external electron source.
\medskip

One of our goals in this paper is to investigate the effects of an energy band gap and temperature on the plasmon
excitation energy dispersion and dissipation in doped monolayer graphene. Specifically, we investigated different
dynamics of graphene electrons for the cases  when either the graphene layer is free-standing (closed system)
or when it is in close proximity with a thick conducting substrate (treated as an electron reservoir
in an open system). The dissipation of plasmons is associated with Landau damping by  particle-hole
modes, which will be modified by the Coulomb coupling of graphene electrons to an ajoining system.
in addition to the shift of the plasmon frequency. Considering the fact that finite temperatures
may also induce plasmon dissipation into particle-hole pairs, we have investigated  how this
reservoir-related dissipation rate depends on temperature.
\medskip

Closed-form analytic expressions for the long-wavelength plasmon dispersion relation in gapped
graphene as well as the dynamical polarization function at zero temperature were reported by
Pyatkovskiy in Ref.\,[\onlinecite{pavlo}]. This important study predicted the existence of dissipation-free plasmons in a large range of wave vector in the presence of a finite energy gap.
Such a bandgap can be opened by either using a substrate or by illuminating graphene with
circularly-polarized light.\,\cite{Kibis} The polarization function for gapped graphene in this
case was derived analytically in Ref.\,[\onlinecite{Busl}].
\medskip

In order to tailor effectively the plasmon frequencies, graphene has been hybridized with
prefabricated plasmonic nanoarrays and metamaterials\,\cite{PR1, PR2, PR3}. Therefore, a
thorough understanding of the dispersion and dissipation of plasmons in graphene interfacing
with different kinds of substrates is necessary for designing innovative devices and
their applications. In this paper, we investigate finite-temperature nonlocal-plasmon dynamics
 in a graphene open system which includes Coulomb-coupling to an electron reservoir (semi-infinite
 conducting substrate). Here, we  concentrate on the dynamics of energy dissipations for both
 surface-plasmon-like (upper) and graphene-like (lower)\,\cite{Poli2, ours}
branches and their energy renormalization by finite temperatures.
\medskip

The rest of the paper is organized as follows. In Sec.\,\ref{sec2}, as an example for the closed
system, we generalized the polarization function for gapless graphene in Ref.\,[\onlinecite{DSmain}]
to one for gapped graphene at finite temperature, in which the analytical expressions for plasmon
dispersion and dissipation rate are derived for both the low- and high-temperature limits. In
Sec.\,\ref{sec3}, we formulate the Coulomb coupling of graphene electrons to an external electron
reservoir. Furthermore, the renormalization of graphene plasmon dispersion, as well as the additional
plasmon  dissipation channel, by interacting with a  reservoir are investigated and associated
numerical results are presented in Sec.\,\ref{sec4}, including effects of finite temperatures and
energy gaps. Finally, some concluding remarks are presented in Sec.\,\ref{sec5}.

\section{Plasmons in gapped-graphene layers at finite temperature}
\label{sec2}

To elucidate clearly the physics of plasmon excitations in proposed gapped-graphene based hybrid
structures, as shown in Fig.\,\ref{FIG:1}, the natural first step in this study is to calculate
analytically  the plasmon modes in gapped graphene at finite temperature by neglecting its interaction
with the  conducting substrate, thereby treating it as a  closed quantum system. In this case,
the plasmon dissipation is entirely determined by the material properties, wave number and temperature, and
the plasmons are determined by the zeros of the dielectric function $\epsilon_{T}(q,\omega;\,\mu)=
1 - V(q)\,\Pi^{(0)}_T (q,\omega; \mu)$. Here $V(q)=2 \pi e^2 / \epsilon_s q$ is the Coulomb interaction
for a two-dimensional (2D) layer. We begin with the non-interacting electron polarizability at zero
temperature, given by\,\cite{W5, Ando, DS07, Wun}

\begin{equation}
\Pi^{(0)}_{T=0}(q,\omega;\,E_F) = -\frac{g g'}{4 \pi^2} \int d^2 {\bf k} \,
f^{s s'} ({\bf q},{\bf k})\, \frac{n_{T=0}[ \varepsilon_s(k)-E_F ] -
n_{T=0}[ \varepsilon_{s'}(\vert {\bf k} + {\bf q} \vert)-E_F ]}{\varepsilon_s(k)
-  \varepsilon_{s'} (\vert {\bf k} + {\bf q} \vert)  + \hbar (\omega + i \gamma)}\ .
\label{pi0}
\end{equation}
Here, we have $g=2$ for spin degeneracy and $g^\prime=2$ for valley degeneracy. Also, $E_F$ is the
Fermi energy at zero temperature, $\varepsilon_s(k) = s \sqrt{\Delta^2+(\hbar v_F k)^2}$ denotes
 the energy dispersion  for the $s=1$ (electron-like) and $s=-1$ (hole-like) subbands,
where $2\Delta$ is the energy gap and  $v_F$ is the Fermi velocity in pristine graphene. The
occupation factors of an electron state
$\vert {\bf q}, s \rangle$ are given by the Fermi-Dirac distribution functions. Finally, the
form factor  $f^{s s'} ({\bf q},{\bf k}) = \vert \langle {\bf q} + {\bf k}, s' \vert
\tet{e}^{i {\bf q} \cdot {\bf r}} \vert {\bf k}, s \rangle \vert^2 $.
\medskip

We now turn to a derivation of  of the finite-temperature polarization function $\Pi^{(0)}_T (q,\omega;\,\mu)$
for gapped graphene. It has been noted\,\cite{15, Yan, DSmain} that the desired polarizability could be obtained as an integral transformation of its corresponding zero-temperature value $\Pi^{(0)}_{T=0}(q,\omega;\,E_F)$:

\begin{equation}
\Pi^{(0)}_T (q,\omega;\,\mu) = \frac{\beta  }{2} \int\limits_{0}^{\infty}  \frac{\Pi^{(0)}_{T=0E} (q,\omega;\,E=E_F)}
{1+ \cosh \left[\beta (\mu - E)\right]}\ ,
\label{mainT}
\end{equation}
where  $\beta=1/k_BT$  and $\mu(T)$ is the electron chemical potential at temperature $T$.
\medskip

An important difference  in the calculation of the polarizability for gapped graphene
for graphene  is that  we must exclude electron energies below the gap $\vert \ve \vert < \Delta$,
which means that we need to introduce another Heaviside step
$\theta(\vert \ve \vert - \Delta)$ besides the actual distribution function. This is essential
 in comparison with gapless graphene or the  2D electron gas. This situation has been considered
 in the calculation of the  zero-temperature polarizability by setting the proper limits for $k$-integrals\,\cite{pavlo}.   One may easily verify that $\theta(\vert \ve \vert - \Delta)$ also accounts for the temperature dependence of the chemical potential $\mu(T)$ as well as the doping dependance at a given temperature used in Eq.\,(\ref{mainT}).

\subsection{Low-Temperature Limit}
\label{sec2.1}

We first consider the low-temperature limit with $k_B T \ll E_F$ (or $k_B T \ll \hbar \omega$
for intrinsic graphene) in the polarization function. For $T \to 0$, the denominator of
Eq.\,\eqref{mainT}  becomes a delta function, scaled as $\beta\tet{e}^{-\beta \vert \mu -E
\vert}$ diverging at $E = \mu$. By assuming that the $\mu$-dependence of the polarization is
smooth for a low enough temperature, one may  expand $\Pi^{(0)}_{T} (q,\omega;\,\mu)$ in a power series. As a result, we arrive at the following expression

\begin{equation}
\Pi^{(0)}_{T} (q,\omega;\,\mu) = \Pi^{(0)}_{T=0} (q,\omega;\,\mu) + \frac{\pi^2}{24} \, (k_B T)^2 \,\frac{\partial^2}{\partial \mu^2} \Pi^{(0)}_{T=0} (q,\omega;\,\mu)\ .
\label{exp}
\end{equation}
We emphasize that the temperature-dependent chemical potential $\mu$ enters  this result
because we expanded the hyperbolic cosine function in Eq.\ (\ref{mainT}) around
$E=\mu$ so that even the first term on the right-hand side of Eq.\ (\ref{exp}) depends
on temperature. We now exploit Eq.\ (\ref{exp}) for gapped graphene in the long wavelength
limit with the help
of\,\cite{pavlo}

\begin{equation}
\Pi^{(0)}_{T=0}(q \ll k_F, \omega;\,\mu) = \frac{\mu}{\pi \hbar^2}
\left[ 1 - \left( \frac{\Delta}{\mu} \right)^2 \right] \,
\frac{q^2}{\omega^2} \ ,
\label{pavloP}
\end{equation}
where $k_F$ is the Fermi wave number for gapped graphene.
This expression is valid only for extrinsic graphene with $E_F > \Delta$. Consequently, the
polarization function becomes

\begin{eqnarray}
&& \Pi^{(0)}_T (q, \omega; \mu) = I(T)\, \frac{q^2}{\omega^2} \\
\nonumber
&& I(T) = \frac{\mu}{\pi \hbar^2} \left[ 1 - \frac{\Delta^2}{\mu^2}  - \frac{\pi^2}{12}
\, \frac{\Delta^2}{\mu^4} (k_B T)^2 \right] \ .
\end{eqnarray}
The corresponding plasmon dispersion relation is $\omega_p^2(q) = (2 \pi e^2 / \epsilon_s) \, I(T) \, q$. We conclude that the plasmon frequency is lowered, in comparison with the result with a bandgap at zero temperature. However, the $\backsimeq q^{1/2}$ dispersive feature is still preserved, as for all 2D materials.
Since the density of states $D(\varepsilon)$ of graphene is unchanged in the presence of an energy gap,
and given by $D(\varepsilon)= 2 \varepsilon / (\pi \hbar^2 v_F^2)$, as well as the chemical potential
$\mu(T)$ at low temperatures ($T \ll T_F$) is calculated as $\mu(T)/E_F \backsimeq 1 - (\pi^2/6)\, (T/T_F)^2$,
the polarization function may be expressed through the zero-temperature Fermi energy as

\begin{equation}
\Pi^{(0)}_T (q, \omega;\,\mu) = \left[\frac{( E_F^2 - \Delta^2 )}{\pi \hbar^2 E_F} - \pi \,\frac{2E_F^2 +3 \Delta^2}{12 \hbar^2 E_F} \, \left(\frac{T}{T_F}\right)^2 \right]\frac{q^2}{\omega^2} \ ,
\end{equation}
where $T_F=E_F/k_B$ is the Fermi temperature.

\subsection{High-Temperature Limit}
\label{sec2.2}

Turning now to the high-temperature limit for the polarization function for gapped graphene,
we still start with the integral transformation in Eq.\,\eqref{mainT} and the zero-temperature
polarization function in Eq.\,\eqref{pavloP} in the long-wavelength limit. These results are
regarded as a generalization of the zero-temperature plasmons in gapped graphene\,\cite{pavlo}
as well as the finite temperature polarization function in graphene with no gap\,\cite{DSmain}.
Our analytical calculations for intrinsic graphene ($\mu=0$) yield (see details in Appendix\ \ref{ap1})

\begin{equation}
\label{PT}
{\rm Re}\,\Pi^{(0)}_T (q,\,\omega;\,\mu=0)= \frac{g_t \ln 2}{2 \pi} \frac{q^2}{\hbar^2 \omega^2}\, k_B T
-\frac{g_t}{4\pi\hbar^2}\,\frac{\Delta^2}{4k_BT}\left[C- \ln\left( \frac{\Delta}{2 k_BT} \right) \right]\frac{q^2}{\omega^2} \ ,
\end{equation}
where $g_t=gg'=4$ is the total degeneracy factor of graphene electrons and $C$ is a small positive constant.
Consequently, we obtain the plasmon dispersion relation as

\begin{equation}
\omega_p^2(q) = \frac{4}{\hbar}\, v_F  r_s q \left\{k_B T \ln 2 -
\frac{\Delta^2}{8 k_B T} \left[C - \ln\left( \frac{\Delta}{2 k_B T} \right)  \right] \right\}\ ,
\label{gap1}
\end{equation}
where $r_s=e^2/\epsilon_s\hbar v_F$ is the so-called graphene fine-structure constant.
\medskip

The dissipation rate $\gamma(q,\,T;\,\mu)$ (inverse lifetime) for the case of weak dissipation is defined as

\begin{equation}
\gamma(q,\,T;\,\mu) = \frac{{\rm Im}\,\Pi^{(0)}_T (q,\omega = \omega_p;\,\mu )}{ \displaystyle{\left.\frac{\partial}{\partial \omega} {\rm Re}\,\Pi^{(0)}_T (q, \omega;\,\mu) \right|_{\omega = \omega_p}  }}\ ,
\label{gammag}
\end{equation}
where the evaluation of the plasmon dispersion $\omega=\omega_p(q)$ is a prerequisite for calculating $\gamma(q,\,T;\,\mu)$.
The imaginary part of the zero-temperature polarization function is given by\,\cite{pavlo}

\begin{equation}
{\rm Im}\,\Pi^{(0)}_{T=0}(q, \omega;\,E_F) = \frac{g_t q^2}{8 \hbar \omega} \left(
1- \frac{1}{2}X_0^2
\right) \theta(\hbar \omega-2 E_F) \ ,
\end{equation}
where

\begin{equation}
X_0 = \sqrt{1+ \frac{4 \Delta^2}{\hbar^2 \left( v_F^2 q^2 - \omega^2 \right)}} \backsimeq
1 - \frac{2 \Delta^2}{\hbar^2 \omega^2}\ ,
\end{equation}
which gives

\begin{equation}
{\rm Im}\,\Pi^{(0)}_{T=0} (q,\omega;\,E_F) = \frac{g_t q^2}{16 \hbar \omega} \left(
1 + \frac{4 \Delta^2}{\hbar^2 \omega^2}
\right) \theta(\hbar \omega-2 E_F ) \ .
\end{equation}
By using Eq.\,\eqref{mainT}, the imaginary part of the finite-temperature polarizability of intrinsic graphene is obtained as

\begin{equation}
{\rm Im}\,\Pi^{(0)}_T (q, \omega;\,\mu=0) = \frac{g_t q^2}{16 \hbar \omega}  \left(
1 + \frac{4 \Delta^2}{\hbar^2 \omega^2}
\right) \int\limits_{\Delta}^{\infty} \frac{d\,\mu'}{4 k_BT} \frac{\theta(\hbar \omega - 2 \mu' )}
{\cosh^2(\mu'/2 k_BT)} = \frac{g_t}{64} \, \frac{q^2}{k_B T} \left( 1 - \frac{\Delta}{\hbar\omega} \right) \ .
\label{impt}
\end{equation}
Consequently, using Eq.\,\eqref{gammag}, we obtain the lowest-order correction $\backsimeq \Delta^2$ to the dissipation rate in intrinsic gapped graphene as

\begin{equation}
\gamma(q,\,T;\,\mu=0) = \frac{\pi}{8} \sqrt{\frac{\hbar}{k_B T}} (\ln 2)^{1/2} (r_s v_F)^{3/2} q^{3/2} - \frac{\pi}{16}\,r_sv_Fq\left(\frac{\Delta}{k_BT}\right) \ .
\end{equation}
From our calculated dissipation rate we find that the existence of a bandgap will slow down the plasmon dissipation.
\medskip

\begin{figure}
\centering
\includegraphics[width=0.55\textwidth]{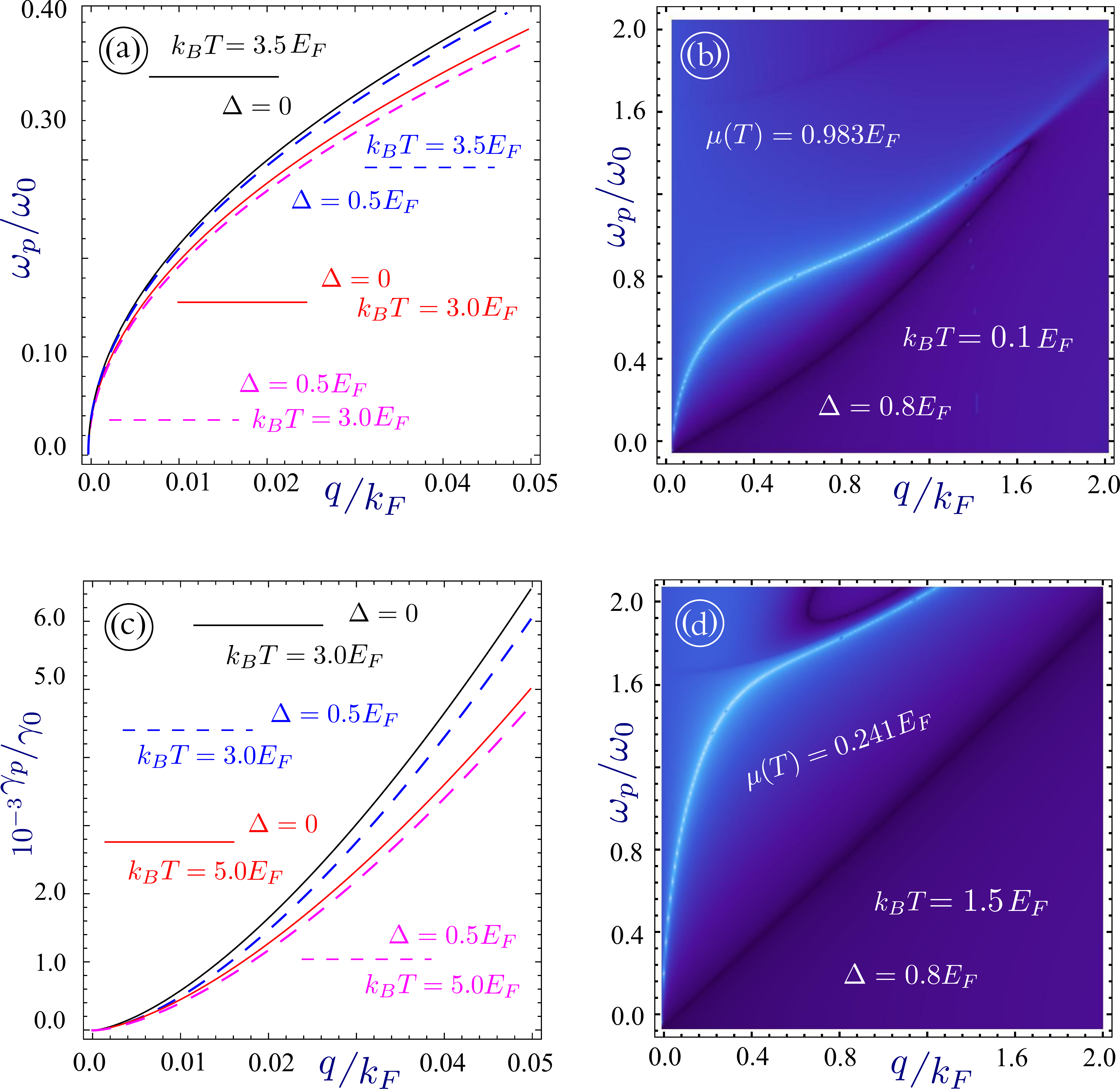}
\caption{(Color online) Analytical (left) and numerical (right) results for plasmons in gapped graphene. Panels $(a)$,
$(c)$ represent respectively the analytical plasmon dispersions and dissipation rates of intrinsic graphene in the long-wavelength limit. Plot $(a)$ shows
the plasmon frequency $\omega_p$ in units of $\omega_0 = (2E_F/\hbar)\sqrt{\ln 2}$,
where $E_F=\hbar v_F k_F  = 0.05\,$eV, while plot $(c)$ displays the dissipation rate $\gamma_p$ in units of $\gamma_0 =(\pi E_F/8\hbar)\sqrt{r_s^3\ln 2}$ for the case of high
temperatures. Different values of $\Delta$ and $k_BT$ are individually labeled in plots $(a)$, $(c)$.
The numerical calculations of plasmon dispersions are presented in panels $(b)$, $(d)$ for two temperatures $k_B T/E_F = 0.1,\, 1.5$ as well as a fixed energy gap $\Delta/E_F = 0.8$.}
\label{FIG:2}
\end{figure}

The plasmon dispersion relation determined  by the change in the density plot peaks with $q$
and the associated dissipation rates proportional to the brightness of these   peaks
 for gapped graphene are presented in Fig.\,\ref{FIG:2} for  chosen values of $T$, $E_F$
 and $\Delta$. The plasmon frequencies $\omega_p$ increase with temperature $T$ in Fig.\,\ref{FIG:2}$(a)$,
 scaled as $\backsimeq \sqrt{T}$ in the long wavelength limit. On the other hand, the
energy gap  slightly reduces $\omega_p$ in the same plot at high temperatures. Therefore,
there is   interplay between thermal and bandgap effects with respect to $\omega_p$.
The dissipation rate $\gamma_p$ of plasmons is determined by the particle-hole mode damping.
As shown in Figs.\,\ref{FIG:2}$(b)$ and \ref{FIG:2}$(d)$, the increase of $T$ leads to an expansion
of the intraband single-particle excitation region below the main diagonal  in the $\omega-q$ space.
Furthermore, the interband particle-hole mode region, above the main diagonal at $k_BT=1.5\,E_F$.
However, in the high-temperature limit  with $(k_BT \gg \hbar v_F q$ and $k_B T \gg \Delta)$,
these interband particle-hole modes are   suppressed, leading to significantly reduced $\gamma_p$ in Fig.\,\ref{FIG:2}$(c)$. Also from Fig.\,\ref{FIG:2}$(c)$, it is interesting to notice the
decrease of $\gamma_p$ with increasing $\Delta$ at intermediate temperatures.
Another distinct feature in Fig.\,\ref{FIG:2}$(d)$ is that for large energy gap
$\Delta/E_F = 0.8$ and $k_BT=1.5\,E_F$, the plasmon mode is kept undamped for a relatively wider
 $q$-range within the gap region between the interband and intraband particle-hole modes.

\section{Plasmons in open graphene-based hybrid systems at finite temperature}
\label{sec3}

\begin{figure}
\centering
\includegraphics[width=0.49\textwidth]{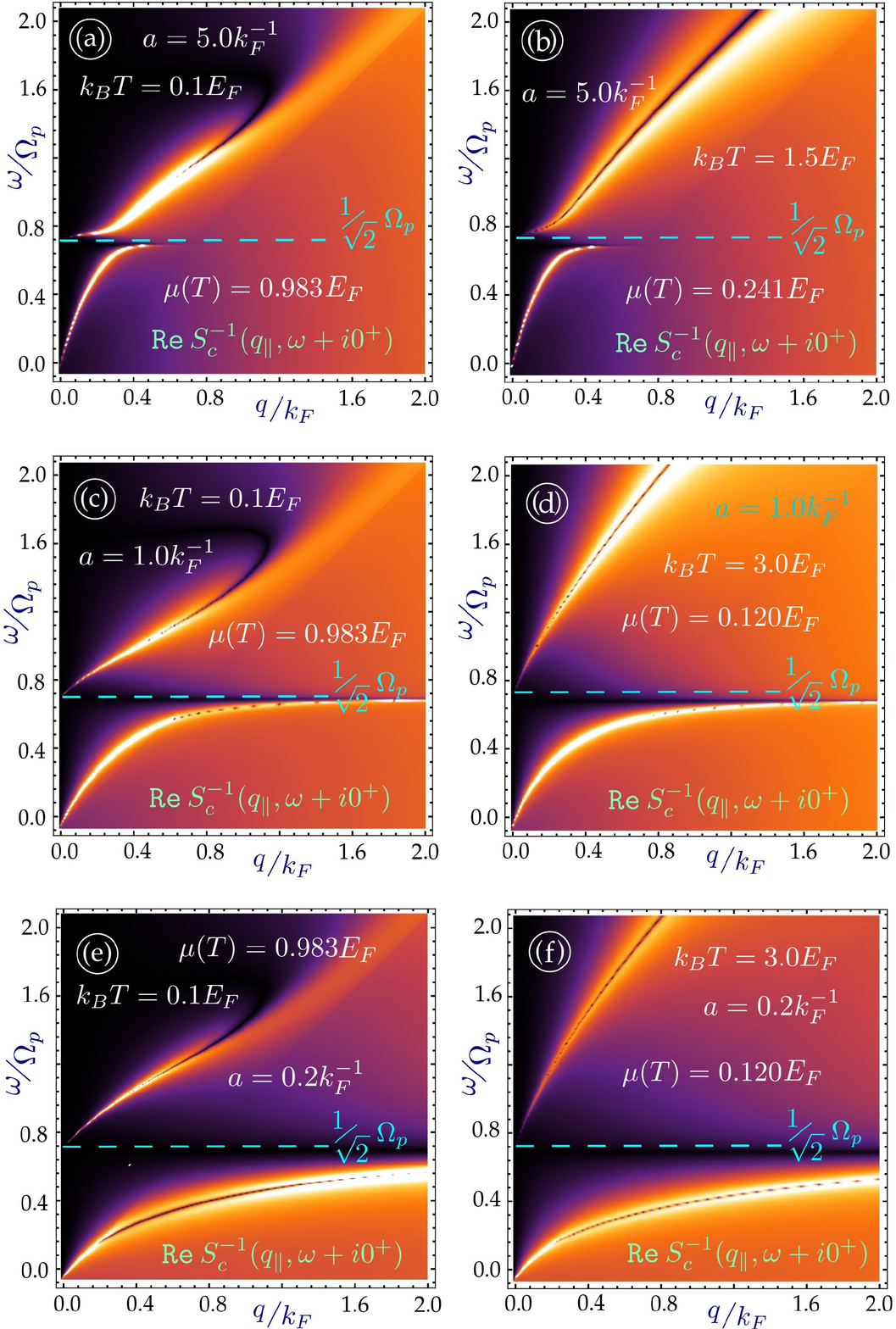}
\caption{(Color online) Density plots of the real part of $S_c(q,\,\omega+i0^+)$ for $\Delta=0$
and doped graphene at chosen temperatures. Peaks of these plots  correspond to the plasmon resonances for
various values of $T$ and $a$. Here, $E_F=\hbar v_Fk_F=10\,$meV, for the electron density
$n=10^{16}\,$m$^{-2}$. Plots in $(a),\,(c),\,(e)$ are for $k_B T = 0.1\,E_F$, while those in
$(b),\,(d),\,(f)$ for $k_B T = 1.5\,E_F$. Additionally, panels $(a),\,(b)$ are for
$a = 5.0\,k_F^{-1}$, but those of $(c),\,(d)$ to $a = 1.0\,k_F^{-1}$. Panels
$(e),\,(f)$ correspond to $a = 0.2\, k_F^{-1}$.}
\label{FIG:3}
\end{figure}

In Sec.\ \ref{sec2}, graphene was  assumed to be embedded in the surrounding host
dielectric material, making  it possible to be treated as a closed quantum system.
However, when the graphene layer is brought close to a conducting medium, the Coulomb
interaction between graphene and the nearby conductor cannot be neglected and the hybrid graphene
system should be regarded as an open system. Physically, graphene is a single atomic-layer material,
which implies that it will inevitably interact with its embedding host  or adjoining conducting
substrate. In considering monolayer graphene  on the surface of a conductor  as an example for open
systems, we introduce additional plasmon dissipation channels due to the Coulomb interaction
between electrons in graphene and the substrate which we model as an electron liquid.
Such an arrangement makes electronic excitation in a graphene layer sensitive to changes in
 their environment, including the presence of a single molecule, electron doping, and even
 thermal fluctuations.
\begin{figure}
\centering
\includegraphics[width=0.49\textwidth]{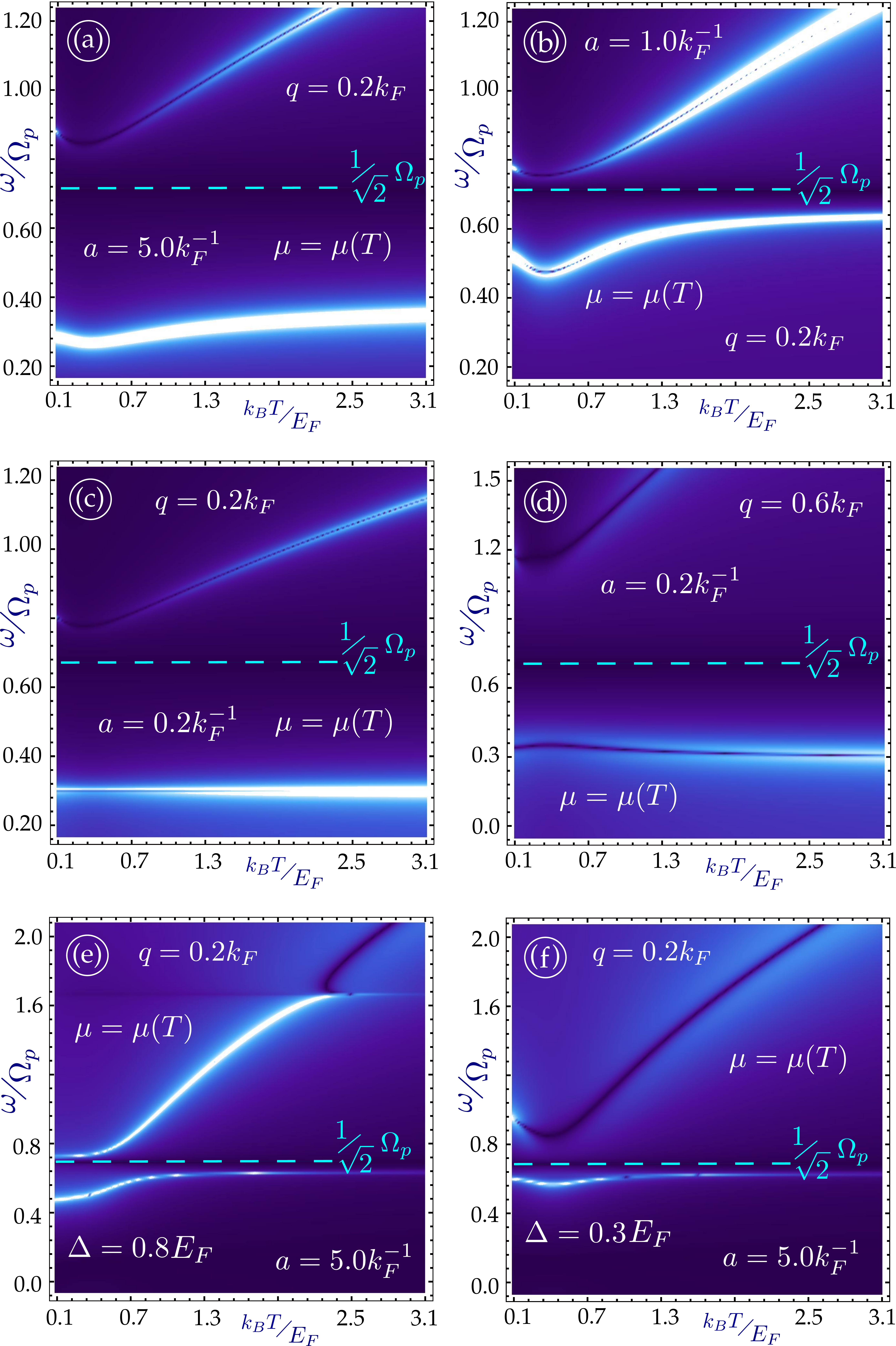}
\caption{(Color online) Density plots of the real part of $S_c(q,\,\omega+i0^+)$ as functions
 of $T$ for doped graphene for two values of $q$. Panels $(a)$-$(d)$ are for $(\Delta = 0)$,
 while panels $(e),\,(f)$ for different energy bandgaps. All plots are shown for $q=0.2\,k_F$
except plot $(d)$ for $q=0.6\,k_F$. In addition, panels $(a),\,(e),\,(f)$ display results for
well-separated graphene layer with $a = 5.0\,k_F^{-1}$, while panels $(b),\,(c)$ for
$a = 0.2\,k_F^{-1}$ and $(d)$ for $a = 1.0\,k_F^{-1}$. Here, $E_F=10\,$meV is chosen.}
\label{FIG:4}
\end{figure}

\subsection{Interaction with Electron Reservoir}
\label{sec3.1}

By including the Coulomb interaction between graphene and the conducting substrate, the nonlocal 
inverse dielectric function of such an open system $\mc{K}(z_1,\,z_2)$ satisfies the following equation
as described in Refs.\ [\onlinecite{Horing2009,ours}]

\begin{equation}
\mc{K}(z_1,\,z_2)= K_S(z_1,\,z_2)+\Pi_T^{(0)}(q,\omega;\,\mu)\,\frac{K_S(a,\,z_2)}{S_{C}(q,\omega;\,\mu)}
\left\{ \int_{-\infty}^\infty dz^\prime\,K_S(z_1,\,z^\prime)\,v_C(q,\,z^\prime-a) \right\}  \ ,
\label{eq:GG-inv}
\end{equation}
where $K_S(z_1,\,z_2)$ denotes the nonlocal inverse dielectric function of the semi-infinite
conducting substrate  with

\begin{eqnarray}
K_S(z,\,z^\prime\vert q,\,\omega) &&  = \, \theta(z)\left\{ \delta(z-z^\prime)+\delta(z^\prime)\,
 e^{-qz} \left[ \frac{1-\epsilon_B(\omega)}{1+\epsilon_B(\omega)} \right]  \right\}
\nonumber\\
                                       && + \, \theta(-z) \left\{ \frac{\delta(z-z^\prime)}{\epsilon_B(\omega)}+\delta(z^\prime)\,
                                         e^{ qz}\, \frac{1}{\epsilon_B(\omega)} \left[ \frac{\epsilon_B
                                        (\omega)-1}{\epsilon_B(\omega)+1} \right]  \right\} \ ,
\end{eqnarray}
and $z>0$ ($z<0$) corresponds to air (conductor) side, respectively, with the surface at $z=0$.
Also,   $\epsilon_B(\omega)=1-\Omega_p^2/\omega^2$ by the Drude model with $\Omega_p$ being
the bulk plasma frequency.
The interaction between graphene and the substrated is included in the second term of Eq.\,\eqref{eq:GG-inv},
$a$ is the distance of the graphene layer from the conducting surface, $v_C(q,\,z-z^\prime)=(2\pi e^2/\epsilon_s)\,\exp(-q|z-z^\prime|)$, and

\begin{equation}
S_{C}(q,\omega;\,\mu)= 1-\frac{2\pi e^2}{\epsilon_s q}\, \Pi_T^{(0)} (q,\omega;\,\mu)
\left\{ 1+ e^{-2qa} \,\frac{1-\epsilon_B(\omega)}{1+\epsilon_B(\omega)}   \right\}\ .
\label{Sc}
\end{equation}
The denominator $S_{C}(q,\omega;\,\mu)$ in Eq.\,\eqref{eq:GG-inv} yields the poles in
the inverse dielectric function and consequently the plasma dispersion relation.

\begin{figure}
\centering
\includegraphics[width=0.45\textwidth]{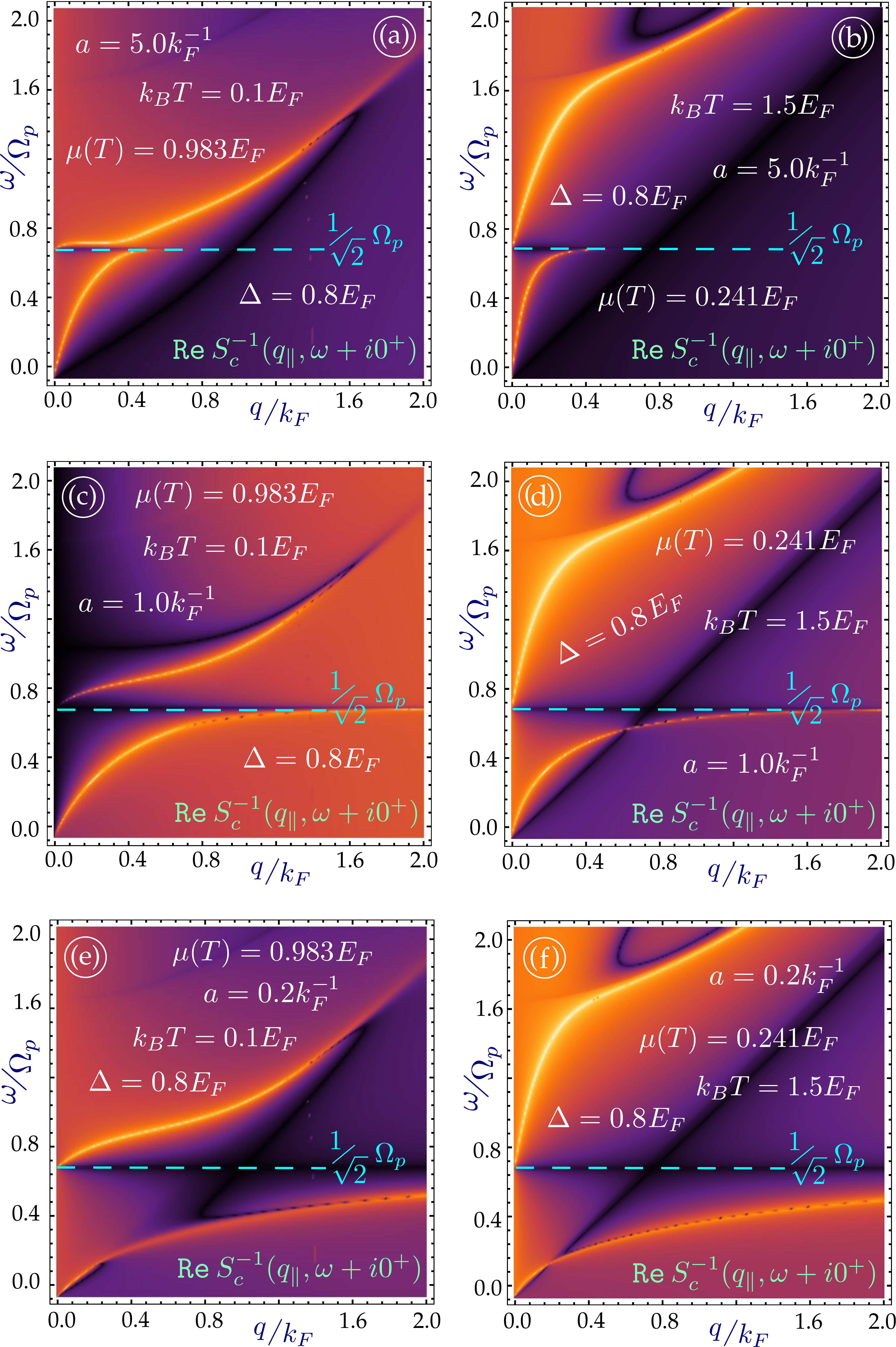}
\caption{(Color online) Density plots of the real part of $S_c(q,\,\omega+i0^+)$ with a large
energy gap $\Delta = 0.8\,E_F$ at finite $T$ for various values of $a$ and $T$. Panels $(a),\,(c),\,(e)$ correspond to a lower temperature with $k_BT=0.1\,E_F$, while panels $(b),\,(d),\,(f)$ to a
higher temperature $k_BT=1.5\,E_F$. Additionally, smaller layer separations are chosen with
$a=5.0\,k_F^{-1}$ in plots $(a),\,(b)$, $a=1.0\,k_F^{-1}$ in plots $(c),\,(d)$ and $a=0.2\,k_F^{-1}$ in plots $(e),\,(f)$, respectively.
Here, $E_F=10\,$meV is taken.}
\label{FIG:5}
\end{figure}

\subsection{Gapless Graphene}
\label{sec3.2}

For open systems, we know from Lindhard's theory that the interaction of a graphene layer
with a conducting substrate  will not only acquire additional interaction for plasmons,
as presented in Eq.\,\eqref{Sc}, but also introduce more dissipation channels for plasmons.
Here, the surface plasmon dissipation into an electron reservoir will be neglected. This is
reasonable  for a large range of wave vectors $q \ll 1/\lambda_F$, where
$\lambda_F \backsimeq 0.5 \,nm$ is the Fermi wavelength in metals comparable with the lattice constant.
It has already been shown that  Landau damping of plasmons plays a crucial role on the high-temperature
dispersion relations. Therefore, we expect the Landau damping of graphene plasmons at high
temperatures will be further modified by its interaction with conducting substrate.
Another factor which may contribute to dissipation includes energy gap opening in a graphene layer.
As reported previously, $\backsim 0.1\,$eV energy gap in graphene could be created not only
by a substrate, but also by irradiation  using circularly-polarized light\,\cite{Kibis}.
Polarization function, plasmons and their dissipation in this case were addressed
in Ref.\ [\onlinecite{Busl}].
\medskip

As an illustration of dissipation physics, we take the graphene polarizability with
$\Delta=0$ in the long-wavelength limit, as given in
Ref.\ [\onlinecite{DSmain}] and Eq.\,\eqref{PT}, and obtain

\begin{equation}
1 - \frac{2 \pi}{q}\, \hbar r_s v_F\, \frac{2
\ln 2}{\pi\hbar^2}\, \frac{q^2}{\omega^2}\, k_BT
\left\{
1 + \tet{e}^{-2qa}\,\frac{\Omega_p^2}{2 \omega^2 - \Omega_p^2}
\right\}=1-q\Lambda_0(T)\left(\frac{\Omega_p}{\omega}\right)^2\left\{
1 + \tet{e}^{-2qa}\,\frac{1}{2 (\omega/\Omega_p)^2 - 1}
\right\} = 0  \ ,
\label{lam}
\end{equation}
where $\Lambda_0(T) = 4 \ln 2\,(r_s v_F/\hbar\Omega_p^2)\, k_B T$ playing the
role of ``effective length'' in the plasmon dispersion $\backsimeq  \sqrt{q\Lambda_0(T)}$.
\medskip

We consider first the case which corresponds to small layer separation or $qa\ll 1$.  For this, we find
in the linear approximation

\begin{eqnarray}
\omega_{p,1}(q) &=& \sqrt{ 8 \ln 2\, \frac{r_s v_F}{\hbar}}\, \sqrt{a}\,\sqrt{k_B T} \, q  \ ,
\nonumber\\
\omega_{p,2}(q) &=& \frac{\Omega_p}{\sqrt{2}} + \sqrt{8} \ln 2 \, \frac{r_s v_F}{\hbar \, \Omega_p}\, k_B T \, q\ .
\end{eqnarray}
Both branches are linear in $q$, and the surface plasmon $\omega_{p2}(q)$ does not depend on
$a$ in this approximation.
\medskip

\begin{figure}
\centering
\includegraphics[width=0.45\textwidth]{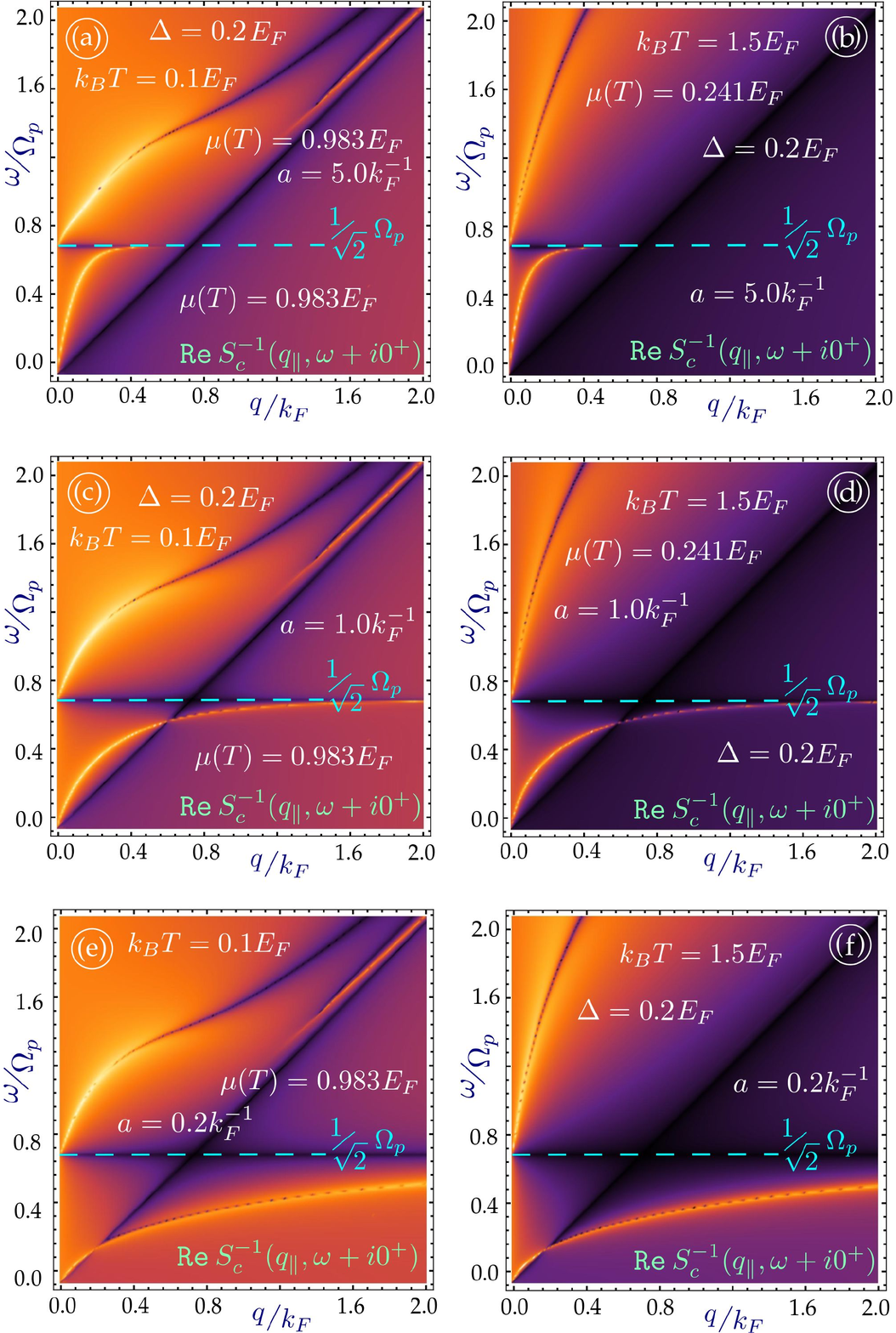}
\caption{(Color online) Density plots of the real part of $S_c(q,\,\omega+i0^+)$ with a large energy gap $\Delta = 0.2\,E_F$ at finite $T$
for various values of $a$ and $T$. Panels $(a),\,(c),\,(e)$ are for a lower temperature $k_BT=0.1\,E_F$, while
panels $(b),\,(d),\,(f)$ for a higher temperature $k_BT=1.5\,E_F$.
Moreover, different layer separations are taken for
$a=5.0\,k_F^{-1}$ in plots $(a),\,(b)$, $a=1.0\,k_F^{-1}$ in plots $(c),\,(d)$ and $a=0.2\,k_F^{-1}$ in plots $(e),\,(f)$, respectively.
Here, $E_F=10\,$meV is chosen.}
\label{FIG:6}
\end{figure}

In the limit when   $qa\gg 1$, a straightforward calculation yields

\begin{eqnarray}
\omega_{p,1}(q)   &=& \Omega_p\sqrt{q\Lambda_0(T)} -\frac{\Omega_p\sqrt{q\Lambda_0(T)}}
{2[1 - 2q \Lambda_0(T)]}\, \tet{e}^{-2qa}  \ ,
\nonumber\\
\omega_{p,2}(q)   &=& \frac{\Omega_p}{\sqrt{2}} + \frac{q\Lambda_0(T)\Omega_p}
{\sqrt{2}[1 - 2 q\Lambda_0(T)]}\,  \tet{e}^{-2qa} \ .
\label{si}
\end{eqnarray}
In addition, within the usual limit of $q\Lambda_0(T) \ll 1$,  we have $1-2q\Lambda_0(T)\approx 1$ and obtain from Eq.\,\eqref{si}

\begin{eqnarray}
\omega_{p,1}(q) &=& \sqrt{\frac{4 \ln 2}{\hbar}}\, \left( v_F r_s q \right)^{1/2}\, \sqrt{k_B T} \left\{
1 - \frac{1}{2}\, \tet{e}^{-2qa}\right\}  \ ,
\nonumber\\
\omega_{p,2}(q) &=& \frac{\Omega_p}{\sqrt{2}} + 4 \ln 2 \, \frac{r_s v_F}{\hbar\Omega_p}\, k_B T \, q \,
\tet{e}^{-2qa} \ ,
\end{eqnarray}
from which follows the isolated plasmon dispersions for graphene and constant $\Omega_p / \sqrt{2}$ surface-plasmon frequency as $a \to \infty$.
\medskip

\begin{figure}
\centering
\includegraphics[width=0.45\textwidth]{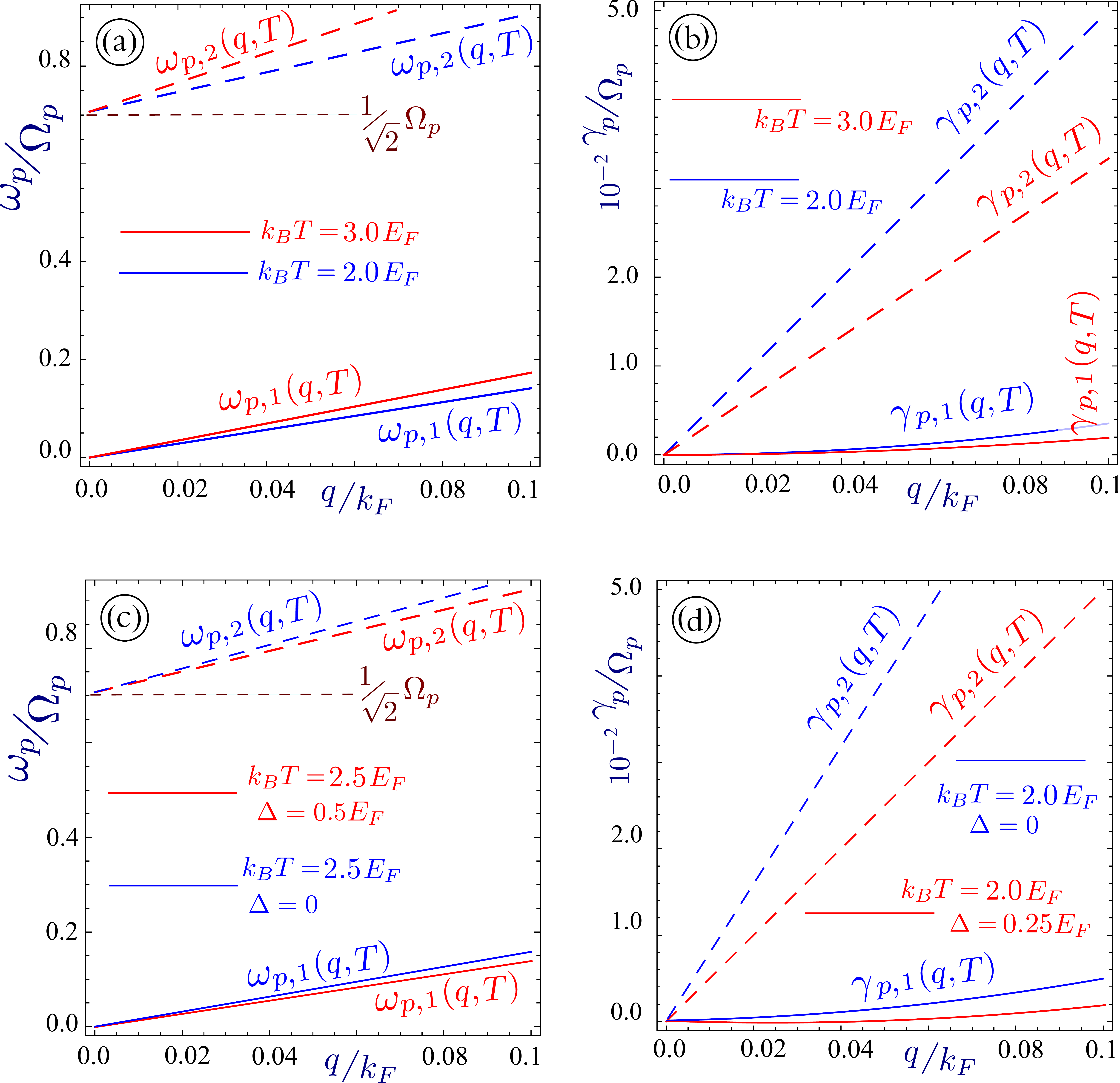}
\caption{(Color online) Plasmon frequencies in panels $(a),\,(c)$) and corresponding
dissipation rates in $(b),\,(d)$) in the long wavelength limit for chosen $T$ and
$\Delta$ values. $\omega_{p,1}$ and $\omega_{p,2}$ in plots $(a),\,(c)$, as well as
$\gamma_{p,1}$ and $\gamma_{p,2}$ in plots $(b),\,(d)$, correspond to lower and upper
plasmon branches. The other parameters, $\Delta$ and $T$, are directly labeled in the figure.
Here, $a=1.0\,k_F^{-1}$ and $E_F=10\,$meV were used  in our calculations.}
\label{FIG:7}
\end{figure}

After computing the plasmon dispersion $\omega_p(q)$, we may use the result  to calculate the
 plasmon dissipation rate $\gamma_p(q)$ for the open system in Fig.\,\ref{FIG:1}.
First, we need to take into account the imaginary part of the polarization function
${\rm Im}\,\Pi^{(0)}_T(q, \omega;\,\mu)$ from the contribution of particle hole modes.
If we only consider small dissipation with $\gamma_p \ll \omega_p$, we can neglect the very small shift of $\omega_p$ from $\backsimeq \gamma_p^2$ and $\backsimeq \gamma_p\,{\rm Im}\,\Pi^{(0)}_T$.
Since Eq.\,\eqref{gammag} is no longer valid after introducing the graphene-conductor Coulomb
coupling in the open system, we have to include both the real and imaginary parts of
${\rm Im}\,\Pi^{(0)}_T(q, \omega;\,\mu)$ in Eq.\,\eqref{lam} and find the dissipation rates
corresponding to each plasmon branch. We know from Eq.\,\eqref{Sc} that
${\rm Im}\,S_{C}(q,\omega;\,\mu) \backsim {\rm Im}\,\Pi_T^{(0)}(q,\omega;\,\mu)$, therefore, the
regions of Landau damping in our open system are determined by the particle-hole modes of a graphene layer.
\medskip

From Eq.\,\eqref{lam}, for $qa\ll 1$ we obtain the dissipation rate for the lower acoustic-like branch

\begin{equation}
\frac{\gamma_{p,1}(q)}{\Omega_p} = \frac{\pi}{4}\, (2 \ln 2)^{1/2} \left( \frac{\hbar}{k_B T}\right)^{1/2}\,\frac{1}{\Omega_p}  \, (r_s v_F)^{3/2} \, a^{3/2} \, q^3 \ ,
\end{equation}
and that for the upper surface-plasmon-like branch

\begin{equation}
\frac{\gamma_{p,2}(q)}{\Omega_p} = \frac{\pi}{16 \sqrt{2}}\, \hbar v_F r_s\, \frac{1}{k_B T}\, q  \ .
\end{equation}
For $qa\gg 1$, on the other hand, the dissipation rate can be approximated as

\begin{eqnarray}
\label{far}
\gamma_{p,1}(q) &=& \frac{\pi}{8}\, \sqrt{\ln 2}\, \sqrt{\frac{\hbar}{k_BT}}\, (r_s v_F)^{3/2} \, q^{3/2}
\left[
1 - 8 \ln 2\, \frac{r_s v_F}{\hbar \Omega_p^2}\, k_BT\, q \,\tet{e}^{-2qa}\right] \ ,
\nonumber\\
\gamma_{p,2}(q) &=&  \frac{\pi}{16}\, \hbar \Omega_p\, \frac{r_s v_F}{k_BT} \, q \,\tet{e}^{-2qa} \ .
\end{eqnarray}

\subsection{Gapped Graphene}
\label{sec3.3}

In the presence of an energy gap for graphene, we expect both the plasmon frequencies and
dissipation rates in the open system will be modified. From our calculation we find that the
results for the plasmon frequencies are similar to those of gapless graphene
except for a modified effective length $\Lambda_\Delta$, i.e.,

\begin{equation}
\Lambda_\Delta(T) =\frac{4}{\hbar \, \Omega_p^2}\, v_F\, r_s \left\{k_BT \ln 2 -
\frac{\Delta^2}{4 k_B T} \left[\mbb{C} - \ln\left( \frac{\Delta}{2 k_BT} \right)  \right] \right\} \ ,
\end{equation}
where $\mbb{C}$ is a small positive constant introduced in Eq.\,\eqref{PT}. Therefore, for
$qa \ll 1$ we obtain

\begin{eqnarray}
&& \omega_{p,1}(q)/\Omega_p = \sqrt{2 a\Lambda_\Delta(T)} \, q\ , \\
\nonumber
&& \omega_{p,2}(q)/\Omega_p = 1/\sqrt{2} + q\Lambda_\Delta(T)/\sqrt{2}\ .
\end{eqnarray}
In the case of $qa \gg 1 $, on the other hand, the solutions are approximated as

\begin{eqnarray}
\omega_{p,1}(q) &=& \Omega_p\sqrt{q\Lambda_\Delta(T)} \left( 1 - \frac{1}{2}\, \tet{e}^{-2qa} \right)\ ,
\nonumber\\
\omega_{p,2}(q) &=& \Omega_p/\sqrt{2} +  \Omega_p\, q\Lambda_\Delta(T)\, \tet{e}^{-2qa}\ .
\end{eqnarray}
The calculations of the dissipation rates in the presence of an energy gap are much more
complicated since they involve the modified imaginary  part of the polarization function
${\rm Im}\,\Pi^{(0)}_T(q, \omega;\,\mu)$. A lengthy calculation for $qa \ll 1$ leads to

\begin{eqnarray}
&& \gamma_{p,1}(q) = \frac{\pi}{4}\, (2 \ln 2)^{1/2} \left( \frac{\hbar}{k_B T}\right)^{1/2} (r_s v_F)^{3/2} \, a^{3/2}\, q^3 - \pi r_sv_Fa\Delta\left(\frac{1}{8k_BT}
- \ln 2\,\frac{r_sv_F}{\hbar\Omega_p^2}\, q\right)q^2\ , \\
\nonumber
&& \gamma_{p,2}(q) = \frac{\pi}{16 \sqrt{2}}\, v_F r_s\, \frac{1}{k_B T} \, \left(\hbar \Omega_p - \sqrt{2}\Delta \right) q  \ .
\end{eqnarray}
We see that $\gamma_{p,1}(q)$ acquires both $\backsimeq q^3$ and $\backsimeq q^2$ terms for a
finite gap, while the linear dispersions for $\gamma_{p,2}(q)$ is preserved.

\section{Numerical Results for Open Systems}
\label{sec4}

Whenever the distance $a$ between the 2D layer and the surface is large, these  two components
are decoupled. Consequently, a pair of  plasmon branches one of which effectively behaves like that
in the absence of the conductor, as shown in Figs.\,\ref{FIG:3}$(a)$ and \ref{FIG:3}$(b)$, except
for a  limited  region of $q$. For $k_B T = 1.5\,E_F$ in these two plots, the upper branch is
significantly damped. In the long wavelength limit, the Coulomb coupling is enhanced. In this case,
as displayed in Figs.\,\ref{FIG:3}$(e)$ and \ref{FIG:3}$(f)$, the ``dominant'' (higher density
peak) plasmon branch is switched to a surface plasmon-like mode approaching
$\Omega_p/\sqrt{2}$ for $q\gg k_F$. This implies an enhancement of graphene-plasmon dissipation
(higher brightness of density plot peak for the lower branch) for small $a$ with an increased
interaction between graphene electrons and the ``external'' electron reservoir in an open system,
as demonstrated by comparing  Fig.\,\ref{FIG:3}$(b)$ with \ref{FIG:3}$(f)$.
Finally, when $k_Fa=1.0$ in Figs.\,\ref{FIG:3}$(c)$ and \ref{FIG:3}$(d)$, these two branches
are strongly coupled to each other with similar density peaks. For a closely located graphene
layer from the surface, the measurement of plasmon dispersion and dissipations was reported in several
recent papers\,\cite{PoliNano, Poli1, Poli2, Poli3, Poli4}. It was discovered that
if the separation becomes small enough, the lower plasmon branch is significantly damped when
it  enters the intraband particle-hole region. However, this plasmon dissipation is lifted up
for large $q$, as seen from Figs.\,\ref{FIG:3}$(e)$ and \ref{FIG:3}$(f)$, due to shrinking of
the intraband particle-hole region.
\medskip

Even for a closed system of monolayer graphene, we are aware that the plasmon dissipation
varies with temperature through the modification of intraband and interband particle-hole
excitations. For  our open system, we expect the graphene-conductor Coulomb coupling will
further change the thermal modulation of  plasmon  dissipation. As an example of the
temperature  dependence, we present in Fig.\,\ref{FIG:4} the plasmon dispersion as a function 
of $T$. For $\Delta=0$ in panels $(a)$-$(d)$, we find the energy of the
 upper branch (surface-plasmon like) exhibits a monotonic increase with $T$, and its dissipation becomes significant for large $q$ and small $a$ values, as seen from Fig.\,\ref{FIG:4}$(d)$.
The lowest dissipation is reached for small $q$ and intermediate $a$ values in Fig.\,\ref{FIG:4}$(b)$.
Interestingly, the lower acoustic like branch remains very low dissipation in nature for
$q=0.2\,k_F$ in Fig.\,\ref{FIG:4}$(a)$-$(c)$ for all temperatures and different values of $a$,
only exhibiting  large dissipation in Fig.\,\ref{FIG:4}(d) for $q=0.6\,k_F$. Temperature dependence
of plasmons with $\Delta\ne 0$ is shown in Fig.\,\ref{FIG:4}$(e)$ and \ref{FIG:4}$(f)$.
Here, a small bandgap in Fig.\,\ref{FIG:4}$(f)$ leads to increased dissipation for the upper 
branch, but this dissipation is suppressed by a large gap in Fig.\,\ref{FIG:4}$(e)$.
\medskip

Numerical results for the open system with a finite energy bandgap in graphene are presented  
in Figs.\,\ref{FIG:5} and \ref{FIG:6} for a large and small gap, respectively. For $(\Delta = 0.8\,E_F$ in Fig.\,\ref{FIG:5}, we find the upper branch remains undamped at large values of $q$ at both low and high
temperatures due to the gap opening between the intraband and interband particle-hole modes. Moreover, 
we observe that increase of $T$ or $\Delta$ gives rise to opposite effects on the shift of plasmon 
energy. It is obvious from Fig.\,\ref{FIG:5} that the energy gap has a profound influence on the
plasmon energy and dissipation for small $q\ll k_F$, where the Coulomb coupling between graphene 
and the conductor becomes strong for fixed layer separation $a$ (scaled as $qa\sim 1$).
Therefore, a smaller separation $a$ implies a larger range of $q$ for significant variation of 
plasmon frequency and dissipation, as shown in Figs.\,\ref{FIG:5}$(e)$ and \ref{FIG:5}$(f)$.
On  the other hand, for small gap $(\Delta = 0.2\,E_F$ in Fig.\,\ref{FIG:6}, the major features in Fig.\,\ref{FIG:3} for the gapless graphene are largely retained. However, change in the upper 
plasmon branch by the energy gap can still be seen especially in the small $q$ region.
\medskip

In order to get a quantitative view of both the plasmon dispersion $\omega_{p,1}(q),\,\omega_{p,2}(q)$ 
and the plasmon dissipation rates $\gamma_{p,1}(q),\,\gamma_{p,2}(q)$,
results for these quantities are presented in Fig.\,\ref{FIG:7} in the long wavelength limit.
From Fig.\,\ref{FIG:7}$(a)$ with $\Delta=0$, we clearly see that  increase in temperature will 
enhance the plasmon energies for both lower and upper branches. On the other hand, for fixed 
$k_BT=2.5\,E_F$ it is found from Fig.\,\ref{FIG:7}$(c)$ that increasing $\Delta$ leads to 
the reduction of linearly $q$-dependent plasmon energies of both branches. Furthermore, we observe 
from Fig.\,\ref{FIG:7}$(b)$ and  $(d)$ that the plasmon dissipation rate of the lower branch ($\sim q^3$) 
is much lower than that of the higher branch ($\sim q$). Here, the dissipation rates of gapless graphene in the open system is reduced by increasing $T$. For gapped graphene in the open system, on the other hand,
the increase of $\Delta$ leads to the enhancement of the plasmon dissipation rates of both branches.

\section{Concluding Remarks}
\label{sec5}

In summary, we have obtained analytic expressions for the plasmon dispersion relations and 
dissipation rates for gapped graphene at high temperature in both closed and open systems.
In the presence of an energy gap $2\Delta$, we have found that the plasmon frequency is 
modified according to $\Delta^2/(4 k_BT) \ln ( \Delta/ (2 k_BT))$, which is different 
from the result at $T=0$. For doped gapped graphene, we have also derived an analytical 
formalism for studying plasmon dispersion and dissipation in the low-temperature limit, 
which was not reported in previous studies.
\medskip

We investigated properties of both intraband and interband particle-hole modes in doped graphene 
and find distinctive behaviors at high temperature for these two types of excitation.
We performed numerical calculations with respect to plasmon dispersion and dissipation for 
an open system  at finite temperature, variable doping and energy gap. When the gap is large, 
a novel feature in the dispersion relation of the upper surface plasmon-like  branch is obtained, 
i.e., opposite shifts of the plasmon frequencies with respect to energy gap and temperature.
For the lower  acoustic-like  branch, as the coupling to the conductor is very strong, dissipation 
is suppressed at large wave vectors due to shrinking of the intraband particle-hole region.
\medskip

Our calculations have shown that  as  the separation between monolayer graphene  and
 the conductor is decreased,  hybridization of the surface and graphene plasmons becomes 
 significant. Additionally, the particle-hole mode contributions  are   
enhanced at the same time. Consequently, strong Landau damping always occurs for these two plasmon 
branches, except when there is  a large energy gap and relatively low temperature. In comparison with the 
gapless graphene, the existence of energy gap reduces the plasmon dissipation, and then, 
stabilizes the plasmon excitations in the open system. Such a linear reduction of dissipation
 with respect to energy gap for the upper plasmon branch, as graphene layer is in proximity 
 with a conductor, is absent for a free-standing graphene layer. Moreover, the dissipation 
 of the lower plasmon branch is essentially suppressed if the energy gap is large.
All of these findings confirm that the coupling to the environment will introduce additional 
dissipation channel for plasmons, and the magnitude of this new dissipation rate varies with 
the energy gap and temperature.

\acknowledgments
This research was supported by  contract \# FA 9453-13-1-0291 of AFRL. DH would 
like to thank the support from the Air Force Office of Scientific Research (AFOSR).

\appendix

\section{Analytical Expression for the Plasmon Dispersion in Gapped Graphene 
at Finite Temperatures}
\label{ap1}

The energy dispersion of gapped graphene may be expressed as 
$\varepsilon(k) = \pm \sqrt{\Delta^2+(\hbar v_F k)^2}$
where $v_F$ is the Fermi velocity and $2\Delta$ is the energy gap 
between the valence (-) and conduction (+) bands.
At zero temperature, the polarization function in the long-wavelength limit is given 
by Eq.\ (\ref{pavloP}) Equation\  (\ref{pavloP})  is valid for $\Delta < \mu$. If $\Delta >\mu$, 
the conduction band is completely empty
and only interband transitions contribute to the plasmon excitations.
We may use Eq.\  (\ref{pavloP}) in Eq.\,\eqref{mainT} with lower limit of integration
$\Delta$ to determine the polarization for finite temperatures. We have
 
\begin{eqnarray}
&&	\Pi^{(0)}_T(q,\omega;\,\mu) = \int\limits_{\Delta}^{\infty} d\,\mu' \frac{\Pi^{(0)}_{T=0}(q,\omega;\,\mu')}{4 k_BT
	\cosh^2\left( \frac{\mu - \mu'}{2 k_B T}
		\right)} = \\
\nonumber				
&&	=	\int\limits_{\Delta}^{\infty} d\,\mu' \
		\frac{\mu'}{\pi \hbar^2}\left[1-\left( \frac{\Delta}{\mu'}\right)^2 \right]
		 \frac{q^2/ \omega^2}{4 k_BT \cosh^2 \left( \frac{\mu - \mu'}{2 k_BT}
		\right)}\ .
\end{eqnarray}
For intrinsic graphene with $\mu=0$, this integral can be split into two parts, i.e.,
$ \Pi^{(0)}_T(q,\omega;\,\mu=0) = I_1+I_2$ with

\begin{eqnarray}
I_1 &=&\int\limits_{\Delta}^{\infty} d\,\mu' \,\frac{\mu'}{\pi \hbar^2}\,
		 \frac{q^2/ \omega^2}{4 k_BT \cosh^2 \left( \mu'/2 k_BT
		\right)} = \frac{2 q^2}{\pi \hbar^2 \omega^2}
\left\{
		\frac{k_BT}{4} \ln 16 - \frac{\Delta}{2}\,\tanh\left(\frac{\Delta}{2 k_BT}\right) + k_BT \ln \left[ \cosh \left(\frac{\Delta}{2 k_BT}\right) \right]
\right\}\ ,		
\nonumber\\
I_2 &=& - \frac{\Delta^2 q^2}{2 k_BT \pi \hbar^2 \omega^2}
\int\limits_{\Delta}^{\infty} \frac{d\, \mu'}{\mu'} \cosh^{-2}\left(\frac{\mu'}{2 k_BT}\right)\ .
\end{eqnarray}
Here, we have considered the fact that, for the case of intrinsic graphene at $T \Longrightarrow 0$, the peak of the integrand occurs at $\mu'=0$, and therefore, an extra factor of $2$ is required.
\medskip

For high temperatures, we approximate

\begin{equation}
I_1 = \frac{1}{\pi \hbar^2}\, \frac{k_BT}{2}\, \frac{q^2}{\omega^2}\, \ln 16
+ \frac{2}{\pi \hbar^2}\, \frac{q^2}{\omega^2}\, G_1(T,\,\Delta) \ ,
\end{equation}
where

\begin{equation}
G_1(T,\,\Delta) = k_BT \ln\left[ \cosh \left(\frac{\Delta}{2 k_BT} \right)\right] -
\frac{\Delta}{2}\,\tanh\left(\frac{\Delta}{2 k_BT}\right) \backsimeq
\frac{\Delta^2}{8 k_BT} - \frac{\Delta^2}{4 k_BT} = - \frac{\Delta^2}{8 k_BT}  \ .
\end{equation}
As a result,  we get

\begin{equation}
 I_1 = \frac{2 \ln 2}{\pi \hbar^2} \, k_BT\, \frac{q^2}{\omega^2} - \frac{1}{\pi \hbar^2} \,\frac{\Delta^2}{4 k_BT} \,\frac{q^2}{\omega^2}\ .
\end{equation}
The  remaining term is rewritten as

\begin{equation}
I_2 = - \frac{1}{2 \pi \hbar^2}\, \frac{\Delta^2}{2 k_BT} \,\frac{q^2}{\omega^2}\, G_2(T, \,\Delta) \ ,
\end{equation}
where

\begin{equation}
G_2(T,\, \Delta) =  \int\limits_{\Delta}^{\infty} \frac{d\mu'}{\mu'}\,
\frac{1}{\cosh^2(\mu'/2 k_BT)}  \ .
\end{equation}
Here, we are looking for an approximated and analytic form for the following integral:

\begin{equation}
 \int\limits_{\delta}^{\infty} \frac{dx}{x \cosh^2 (x)}  \ ,
\end{equation}
where $\delta=\Delta/k_BT\ll 1$. In order to avoid the singularity appearing as
$\delta \to 0$, we perform the integration by parts. Consequently, we obtain

\begin{equation}
 \int\limits_{\delta}^{\infty} \frac{dx}{x\, \cosh^2 x} =  \frac{\ln x}{\cosh^2(x)} \Bigg|_{\delta}^{\infty} +
 2 \int\limits_{\delta}^{\infty} dx\   \frac{\ln x\, \tanh (x) \, }{\cosh^2 (x)}  \ .
\end{equation}
For $\delta \to 0$ or $\Delta \ll k_BT$, the first term gives

\begin{equation}
 \frac{\ln x}{\cosh^2 (x)} \Bigg|_{\delta}^{\infty} = - \ln \left( \frac{\Delta}{2 k_BT}  \right)\ .
\end{equation}
The second part is  a small correction since the integrand does not diverge for $\delta \to 0$, i.e.,

\begin{equation}
 \mbb{C}_1 = \int\limits_{\delta}^{\infty} dx\, \frac{\ln x\, \tanh (x) \, }{\cosh^2 (x)} \backsimeq \int\limits_{0}^{\infty}  dx\,
 \frac{\ln x\, \tanh (x) \,  }{\cosh^2 (x)} = \frac{1}{6} \left[ 36\, \ln \mc{A}_{GK} - 7 \ln 2- 3 (1 + \gamma_{Eu}) \right]  \, ,
\end{equation}
where $\mc{A}_{GK} \backsimeq 1.2824$ is the  Glaisher-Kinkelin constant, defined by

\begin{equation}
 \mc{A}  = \left\{ \left( \prod_{s=1}^{\nu - 1} s^{s} \right) \nu^{-\nu^2/2-\nu/2-1/12} \tet{e}^{\nu^2/4}
 \right\}_{\nu \rightarrow \infty}
\end{equation}
and $\gamma_{Eu} = 0.5772$ is the Euler-Mascheroni constant, given by

\begin{equation}
 \gamma_{Eu} = \left\{ \sum_{s = 1}^{\nu} \frac{1}{s} -\ln \nu \right\}_{\nu \rightarrow 0}\ .
\end{equation}
To make the evaluation more accurate, the numerical integration gives

\begin{equation}
 \int\limits_{0}^{\infty} dx\, \frac{\ln x\, \tanh (x) }{\cosh^2 (x)} = -0.1048\ ,
\end{equation}
and for $\delta = \Delta / (2 k_BT) = 0.05$, we have

\begin{equation}
\int\limits_{\delta}^{\infty} dx\, \frac{\ln x\, \tanh (x) }{\cosh^2 (x)} = -0.1004 \ .
\end{equation}
Eventually, we arrive at the result

\begin{equation}
  G_2 (T,\, \Delta) = \int\limits_{\delta \ll 1}^{\infty} \frac{dx}{x\, \cosh^2 (x)} \backsimeq - \ln\left( \frac{\Delta}{2 k_BT} \right) + \mbb{C}_2 > 0
\end{equation}
even though $\mbb{C}_2 = 2 \mbb{C}_1 \backsimeq -0.20 < 0$.
\medskip

Finally, the polarization function take the form

\begin{equation}
\Pi^{(0)}_T (q,\omega;\,\mu=0) = I_1 + I_2 = \frac{2 \ln 2}{\pi \hbar^2}\,  k_BT\, \frac{q^2}{\omega^2} - \frac{1}{\pi \hbar^2}\, \frac{\Delta^2}{4 k_BT}  \left[  - \ln\left( \frac{\Delta}{2 k_BT} \right) + \mbb{C} \right] \frac{q^2}{\omega^2}\ ,
\end{equation}
where $ \mbb{C} = 1 +  \mbb{C}_2 \approx 0.8$.
The  finite-temperature dielectric function $\epsilon_T (q,\,\omega)$ for intrinsic gapped graphene is

\begin{equation}
\epsilon_T (q,\, \omega) = 1 - \frac{2 \pi e^2}{\epsilon_s q}\, \Pi^{(0)}_T (q, \omega;\,\mu=0) = 1 -
 \frac{2 \pi}{q}\,  r_s \hbar v_F  \,\Pi^{(0)}_T (q, \omega;\,\mu=0) \ .
\end{equation}
From this, we deduce the temperature-dependent plasma frequency of gapped graphene given by

\begin{equation}
\omega^2_p(q) = \frac{4}{\hbar}\, v_F r_s q\left\{k_BT \ln 2 -
\frac{\Delta^2}{8 k_BT} \left[\mbb{C} - \ln\left( \frac{\Delta}{2 k_BT} \right)  \right] \right\}\ .
\label{GG1}
\end{equation}

\bibliography{TempB}

\begin{thebibliography}{53}
\expandafter\ifx\csname natexlab\endcsname\relax\def\natexlab#1{#1}\fi
\expandafter\ifx\csname bibnamefont\endcsname\relax
  \def\bibnamefont#1{#1}\fi
\expandafter\ifx\csname bibfnamefont\endcsname\relax
  \def\bibfnamefont#1{#1}\fi
\expandafter\ifx\csname citenamefont\endcsname\relax
  \def\citenamefont#1{#1}\fi
\expandafter\ifx\csname url\endcsname\relax
  \def\url#1{\texttt{#1}}\fi
\expandafter\ifx\csname urlprefix\endcsname\relax\def\urlprefix{URL }\fi
\providecommand{\bibinfo}[2]{#2}
\providecommand{\eprint}[2][]{\url{#2}}

\bibitem[{\citenamefont{Wunsch et~al.}(2006)\citenamefont{Wunsch, Stauber,
  Sols, and Guinea}}]{Wun}
\bibinfo{author}{\bibfnamefont{B.}~\bibnamefont{Wunsch}},
  \bibinfo{author}{\bibfnamefont{T.}~\bibnamefont{Stauber}},
  \bibinfo{author}{\bibfnamefont{F.}~\bibnamefont{Sols}}, \bibnamefont{and}
  \bibinfo{author}{\bibfnamefont{F.}~\bibnamefont{Guinea}},
  \bibinfo{journal}{New Journal of Physics} \textbf{\bibinfo{volume}{8}},
  \bibinfo{pages}{318} (\bibinfo{year}{2006}).

\bibitem[{\citenamefont{Hwang and Sarma}(2007)}]{DS07}
\bibinfo{author}{\bibfnamefont{E.}~\bibnamefont{Hwang}} \bibnamefont{and}
  \bibinfo{author}{\bibfnamefont{S.~D.} \bibnamefont{Sarma}},
  \bibinfo{journal}{Physical Review B} \textbf{\bibinfo{volume}{75}},
  \bibinfo{pages}{205418} (\bibinfo{year}{2007}).

\bibitem[{\citenamefont{Pyatkovskiy}(2009)}]{pavlo}
\bibinfo{author}{\bibfnamefont{P.}~\bibnamefont{Pyatkovskiy}},
  \bibinfo{journal}{Journal of Physics: Condensed Matter}
  \textbf{\bibinfo{volume}{21}}, \bibinfo{pages}{025506}
  (\bibinfo{year}{2009}).

\bibitem[{\citenamefont{Sarma and Li}(2013)}]{DSmain}
\bibinfo{author}{\bibfnamefont{S.~D.} \bibnamefont{Sarma}} \bibnamefont{and}
  \bibinfo{author}{\bibfnamefont{Q.}~\bibnamefont{Li}},
  \bibinfo{journal}{Physical Review B} \textbf{\bibinfo{volume}{87}},
  \bibinfo{pages}{235418} (\bibinfo{year}{2013}).

\bibitem[{\citenamefont{Abergel et~al.}(2010)\citenamefont{Abergel, Apalkov,
  Berashevich, Ziegler, and Chakraborty}}]{Chakraborty}
\bibinfo{author}{\bibfnamefont{D.}~\bibnamefont{Abergel}},
  \bibinfo{author}{\bibfnamefont{V.}~\bibnamefont{Apalkov}},
  \bibinfo{author}{\bibfnamefont{J.}~\bibnamefont{Berashevich}},
  \bibinfo{author}{\bibfnamefont{K.}~\bibnamefont{Ziegler}}, \bibnamefont{and}
  \bibinfo{author}{\bibfnamefont{T.}~\bibnamefont{Chakraborty}},
  \bibinfo{journal}{Advances in Physics} \textbf{\bibinfo{volume}{59}},
  \bibinfo{pages}{261} (\bibinfo{year}{2010}).

\bibitem[{\citenamefont{Godfrey~Gumbs}(2011)}]{Gb}
\bibinfo{author}{\bibfnamefont{D.~H.} \bibnamefont{Godfrey~Gumbs}},
  \emph{\bibinfo{title}{Properties of Interacting Low-Dimensional Systems}}
  (\bibinfo{publisher}{Wiley-VCH}, \bibinfo{year}{2011}),
  \bibinfo{edition}{isbn: 978-3-527-40894-8} ed.

\bibitem[{\citenamefont{Rold{\'a}n et~al.}(2009)\citenamefont{Rold{\'a}n,
  Fuchs, and Goerbig}}]{Rafael}
\bibinfo{author}{\bibfnamefont{R.}~\bibnamefont{Rold{\'a}n}},
  \bibinfo{author}{\bibfnamefont{J.-N.} \bibnamefont{Fuchs}}, \bibnamefont{and}
  \bibinfo{author}{\bibfnamefont{M.}~\bibnamefont{Goerbig}},
  \bibinfo{journal}{Physical Review B} \textbf{\bibinfo{volume}{80}},
  \bibinfo{pages}{085408} (\bibinfo{year}{2009}).

\bibitem[{\citenamefont{Politano and Chiarello}(2014)}]{PoliNano}
\bibinfo{author}{\bibfnamefont{A.}~\bibnamefont{Politano}} \bibnamefont{and}
  \bibinfo{author}{\bibfnamefont{G.}~\bibnamefont{Chiarello}},
  \bibinfo{journal}{Nanoscale} \textbf{\bibinfo{volume}{6}},
  \bibinfo{pages}{10927} (\bibinfo{year}{2014}).

\bibitem[{\citenamefont{Politano and Chiarello}(2015)}]{Poli1}
\bibinfo{author}{\bibfnamefont{A.}~\bibnamefont{Politano}} \bibnamefont{and}
  \bibinfo{author}{\bibfnamefont{G.}~\bibnamefont{Chiarello}},
  \bibinfo{journal}{Progress in Surface Science}  (\bibinfo{year}{2015}).

\bibitem[{\citenamefont{Politano et~al.}(2012)\citenamefont{Politano, Marino,
  and Chiarello}}]{Poli2}
\bibinfo{author}{\bibfnamefont{A.}~\bibnamefont{Politano}},
  \bibinfo{author}{\bibfnamefont{A.~R.} \bibnamefont{Marino}},
  \bibnamefont{and}
  \bibinfo{author}{\bibfnamefont{G.}~\bibnamefont{Chiarello}},
  \bibinfo{journal}{Phys. Rev. B} \textbf{\bibinfo{volume}{86}},
  \bibinfo{pages}{085420} (\bibinfo{year}{2012}).

\bibitem[{\citenamefont{Politano et~al.}(2011)\citenamefont{Politano, Marino,
  Formoso, Far\'{i}as, Miranda, and Chiarello}}]{Poli3}
\bibinfo{author}{\bibfnamefont{A.}~\bibnamefont{Politano}},
  \bibinfo{author}{\bibfnamefont{A.~R.} \bibnamefont{Marino}},
  \bibinfo{author}{\bibfnamefont{V.}~\bibnamefont{Formoso}},
  \bibinfo{author}{\bibfnamefont{D.}~\bibnamefont{Far\'{i}as}},
  \bibinfo{author}{\bibfnamefont{R.}~\bibnamefont{Miranda}}, \bibnamefont{and}
  \bibinfo{author}{\bibfnamefont{G.}~\bibnamefont{Chiarello}},
  \bibinfo{journal}{Phys. Rev. B} \textbf{\bibinfo{volume}{84}},
  \bibinfo{pages}{033401} (\bibinfo{year}{2011}).

\bibitem[{\citenamefont{Politano et~al.}(2013)\citenamefont{Politano, Formoso,
  and Chiarello}}]{Poli4}
\bibinfo{author}{\bibfnamefont{A.}~\bibnamefont{Politano}},
  \bibinfo{author}{\bibfnamefont{V.}~\bibnamefont{Formoso}}, \bibnamefont{and}
  \bibinfo{author}{\bibfnamefont{G.}~\bibnamefont{Chiarello}},
  \bibinfo{journal}{Journal of Physics: Condensed Matter}
  \textbf{\bibinfo{volume}{25}}, \bibinfo{pages}{345303}
  (\bibinfo{year}{2013}).

\bibitem[{\citenamefont{Zayats et~al.}(2005)\citenamefont{Zayats, Smolyaninov,
  and Maradudin}}]{RG1}
\bibinfo{author}{\bibfnamefont{A.~V.} \bibnamefont{Zayats}},
  \bibinfo{author}{\bibfnamefont{I.~I.} \bibnamefont{Smolyaninov}},
  \bibnamefont{and} \bibinfo{author}{\bibfnamefont{A.~A.}
  \bibnamefont{Maradudin}}, \bibinfo{journal}{Physics reports}
  \textbf{\bibinfo{volume}{408}}, \bibinfo{pages}{131} (\bibinfo{year}{2005}).

\bibitem[{\citenamefont{Halpern et~al.}(2013)\citenamefont{Halpern, Wood, Wang,
  and Corn}}]{RG2}
\bibinfo{author}{\bibfnamefont{A.~R.} \bibnamefont{Halpern}},
  \bibinfo{author}{\bibfnamefont{J.~B.} \bibnamefont{Wood}},
  \bibinfo{author}{\bibfnamefont{Y.}~\bibnamefont{Wang}}, \bibnamefont{and}
  \bibinfo{author}{\bibfnamefont{R.~M.} \bibnamefont{Corn}},
  \bibinfo{journal}{ACS nano} \textbf{\bibinfo{volume}{8}},
  \bibinfo{pages}{1022} (\bibinfo{year}{2013}).

\bibitem[{\citenamefont{Srituravanich et~al.}(2004)\citenamefont{Srituravanich,
  Fang, Sun, Luo, and Zhang}}]{RG3}
\bibinfo{author}{\bibfnamefont{W.}~\bibnamefont{Srituravanich}},
  \bibinfo{author}{\bibfnamefont{N.}~\bibnamefont{Fang}},
  \bibinfo{author}{\bibfnamefont{C.}~\bibnamefont{Sun}},
  \bibinfo{author}{\bibfnamefont{Q.}~\bibnamefont{Luo}}, \bibnamefont{and}
  \bibinfo{author}{\bibfnamefont{X.}~\bibnamefont{Zhang}},
  \bibinfo{journal}{Nano letters} \textbf{\bibinfo{volume}{4}},
  \bibinfo{pages}{1085} (\bibinfo{year}{2004}).

\bibitem[{\citenamefont{Nordlander et~al.}(2004)\citenamefont{Nordlander,
  Oubre, Prodan, Li, and Stockman}}]{Nord1}
\bibinfo{author}{\bibfnamefont{P.}~\bibnamefont{Nordlander}},
  \bibinfo{author}{\bibfnamefont{C.}~\bibnamefont{Oubre}},
  \bibinfo{author}{\bibfnamefont{E.}~\bibnamefont{Prodan}},
  \bibinfo{author}{\bibfnamefont{K.}~\bibnamefont{Li}}, \bibnamefont{and}
  \bibinfo{author}{\bibfnamefont{M.~I.} \bibnamefont{Stockman}},
  \bibinfo{journal}{Nano Letters} \textbf{\bibinfo{volume}{4}},
  \bibinfo{pages}{899} (\bibinfo{year}{2004}).

\bibitem[{\citenamefont{Gumbs et~al.}(2014{\natexlab{a}})\citenamefont{Gumbs,
  Balassis, Iurov, and Fekete}}]{FO1}
\bibinfo{author}{\bibfnamefont{G.}~\bibnamefont{Gumbs}},
  \bibinfo{author}{\bibfnamefont{A.}~\bibnamefont{Balassis}},
  \bibinfo{author}{\bibfnamefont{A.}~\bibnamefont{Iurov}}, \bibnamefont{and}
  \bibinfo{author}{\bibfnamefont{P.}~\bibnamefont{Fekete}},
  \bibinfo{journal}{The Scientific World Journal}
  \textbf{\bibinfo{volume}{2014}} (\bibinfo{year}{2014}{\natexlab{a}}).

\bibitem[{\citenamefont{Gumbs et~al.}(2014{\natexlab{b}})\citenamefont{Gumbs,
  Iurov, Balassis, and Huang}}]{FO2}
\bibinfo{author}{\bibfnamefont{G.}~\bibnamefont{Gumbs}},
  \bibinfo{author}{\bibfnamefont{A.}~\bibnamefont{Iurov}},
  \bibinfo{author}{\bibfnamefont{A.}~\bibnamefont{Balassis}}, \bibnamefont{and}
  \bibinfo{author}{\bibfnamefont{D.}~\bibnamefont{Huang}},
  \bibinfo{journal}{Journal of Physics: Condensed Matter}
  \textbf{\bibinfo{volume}{26}}, \bibinfo{pages}{135601}
  (\bibinfo{year}{2014}{\natexlab{b}}).

\bibitem[{\citenamefont{Iurov et~al.}(2014)\citenamefont{Iurov, Gumbs, Gao, and
  Huang}}]{FO3}
\bibinfo{author}{\bibfnamefont{A.}~\bibnamefont{Iurov}},
  \bibinfo{author}{\bibfnamefont{G.}~\bibnamefont{Gumbs}},
  \bibinfo{author}{\bibfnamefont{B.}~\bibnamefont{Gao}}, \bibnamefont{and}
  \bibinfo{author}{\bibfnamefont{D.}~\bibnamefont{Huang}},
  \bibinfo{journal}{Applied Physics Letters} \textbf{\bibinfo{volume}{104}},
  \bibinfo{pages}{203103} (\bibinfo{year}{2014}).

\bibitem[{\citenamefont{Balassis and Gumbs}(2014)}]{Anto}
\bibinfo{author}{\bibfnamefont{A.}~\bibnamefont{Balassis}} \bibnamefont{and}
  \bibinfo{author}{\bibfnamefont{G.}~\bibnamefont{Gumbs}},
  \bibinfo{journal}{Phys. Rev. B} \textbf{\bibinfo{volume}{90}},
  \bibinfo{pages}{075431} (\bibinfo{year}{2014}).

\bibitem[{\citenamefont{Zhemchuzhna et~al.}(2015)\citenamefont{Zhemchuzhna,
  Gumbs, Iurov, Huang, and Gao}}]{FOLiu}
\bibinfo{author}{\bibfnamefont{L.}~\bibnamefont{Zhemchuzhna}},
  \bibinfo{author}{\bibfnamefont{G.}~\bibnamefont{Gumbs}},
  \bibinfo{author}{\bibfnamefont{A.}~\bibnamefont{Iurov}},
  \bibinfo{author}{\bibfnamefont{D.}~\bibnamefont{Huang}}, \bibnamefont{and}
  \bibinfo{author}{\bibfnamefont{B.}~\bibnamefont{Gao}},
  \bibinfo{journal}{Physics of Plasmas} \textbf{\bibinfo{volume}{22}},
  \bibinfo{eid}{032116} (\bibinfo{year}{2015}).

\bibitem[{\citenamefont{Dresselhaus et~al.}(1996)\citenamefont{Dresselhaus,
  Dresselhaus, and Eklund}}]{nt1}
\bibinfo{author}{\bibfnamefont{M.~S.} \bibnamefont{Dresselhaus}},
  \bibinfo{author}{\bibfnamefont{G.}~\bibnamefont{Dresselhaus}},
  \bibnamefont{and} \bibinfo{author}{\bibfnamefont{P.~C.}
  \bibnamefont{Eklund}}, \emph{\bibinfo{title}{Science of Fullerenes and Carbon
  Nanotubes: Their Properties and Applications}} (\bibinfo{publisher}{Academic
  Press}, \bibinfo{year}{1996}), \bibinfo{edition}{isbn-13: 978-0122218200} ed.

\bibitem[{\citenamefont{Lin and Shung}(1994)}]{nt2}
\bibinfo{author}{\bibfnamefont{M.~F.} \bibnamefont{Lin}} \bibnamefont{and}
  \bibinfo{author}{\bibfnamefont{K.~W.~K.} \bibnamefont{Shung}},
  \bibinfo{journal}{Phys. Rev. B} \textbf{\bibinfo{volume}{50}},
  \bibinfo{pages}{17744} (\bibinfo{year}{1994}).

\bibitem[{\citenamefont{Liu et~al.}(2011)\citenamefont{Liu, Feng, and
  Yao}}]{silicene}
\bibinfo{author}{\bibfnamefont{C.-C.} \bibnamefont{Liu}},
  \bibinfo{author}{\bibfnamefont{W.}~\bibnamefont{Feng}}, \bibnamefont{and}
  \bibinfo{author}{\bibfnamefont{Y.}~\bibnamefont{Yao}},
  \bibinfo{journal}{Physical review letters} \textbf{\bibinfo{volume}{107}},
  \bibinfo{pages}{076802} (\bibinfo{year}{2011}).

\bibitem[{\citenamefont{Tabert and Nicol}(2014)}]{SP}
\bibinfo{author}{\bibfnamefont{C.~J.} \bibnamefont{Tabert}} \bibnamefont{and}
  \bibinfo{author}{\bibfnamefont{E.~J.} \bibnamefont{Nicol}},
  \bibinfo{journal}{Physical Review B} \textbf{\bibinfo{volume}{89}},
  \bibinfo{pages}{195410} (\bibinfo{year}{2014}).

\bibitem[{\citenamefont{Campisi et~al.}(2011)\citenamefont{Campisi, H\"anggi,
  and Talkner}}]{dhh-2}
\bibinfo{author}{\bibfnamefont{M.}~\bibnamefont{Campisi}},
  \bibinfo{author}{\bibfnamefont{P.}~\bibnamefont{H\"anggi}}, \bibnamefont{and}
  \bibinfo{author}{\bibfnamefont{P.}~\bibnamefont{Talkner}},
  \bibinfo{journal}{Rev. Mod. Phys.} \textbf{\bibinfo{volume}{83}},
  \bibinfo{pages}{771} (\bibinfo{year}{2011}).

\bibitem[{\citenamefont{Campisi et~al.}(2009)\citenamefont{Campisi, Talkner,
  and H\"anggi}}]{dhh-3}
\bibinfo{author}{\bibfnamefont{M.}~\bibnamefont{Campisi}},
  \bibinfo{author}{\bibfnamefont{P.}~\bibnamefont{Talkner}}, \bibnamefont{and}
  \bibinfo{author}{\bibfnamefont{P.}~\bibnamefont{H\"anggi}},
  \bibinfo{journal}{Phys. Rev. Lett.} \textbf{\bibinfo{volume}{102}},
  \bibinfo{pages}{210401} (\bibinfo{year}{2009}).

\bibitem[{\citenamefont{Esposito et~al.}(2009)\citenamefont{Esposito, Harbola,
  and Mukamel}}]{dhh-4}
\bibinfo{author}{\bibfnamefont{M.}~\bibnamefont{Esposito}},
  \bibinfo{author}{\bibfnamefont{U.}~\bibnamefont{Harbola}}, \bibnamefont{and}
  \bibinfo{author}{\bibfnamefont{S.}~\bibnamefont{Mukamel}},
  \bibinfo{journal}{Rev. Mod. Phys.} \textbf{\bibinfo{volume}{81}},
  \bibinfo{pages}{1665} (\bibinfo{year}{2009}).

\bibitem[{\citenamefont{Crooks}(2008)}]{dhh-5}
\bibinfo{author}{\bibfnamefont{G.~E.} \bibnamefont{Crooks}},
  \bibinfo{journal}{Journal of Statistical Mechanics: Theory and Experiment}
  \textbf{\bibinfo{volume}{2008}}, \bibinfo{pages}{10023}
  (\bibinfo{year}{2008}).

\bibitem[{\citenamefont{Mukamel}(2003)}]{dhh-6}
\bibinfo{author}{\bibfnamefont{S.}~\bibnamefont{Mukamel}},
  \bibinfo{journal}{Phys. Rev. Lett.} \textbf{\bibinfo{volume}{90}},
  \bibinfo{pages}{170604} (\bibinfo{year}{2003}).

\bibitem[{\citenamefont{De~Roeck and Maes}(2004)}]{dhh-7}
\bibinfo{author}{\bibfnamefont{W.}~\bibnamefont{De~Roeck}} \bibnamefont{and}
  \bibinfo{author}{\bibfnamefont{C.}~\bibnamefont{Maes}},
  \bibinfo{journal}{Phys. Rev. E} \textbf{\bibinfo{volume}{69}},
  \bibinfo{pages}{026115} (\bibinfo{year}{2004}).

\bibitem[{\citenamefont{Crooks}(1999)}]{dhh-8}
\bibinfo{author}{\bibfnamefont{G.~E.} \bibnamefont{Crooks}},
  \bibinfo{journal}{Phys. Rev. E} \textbf{\bibinfo{volume}{60}},
  \bibinfo{pages}{2721} (\bibinfo{year}{1999}).

\bibitem[{\citenamefont{Setiawan and Sarma}(2015)}]{Rec}
\bibinfo{author}{\bibfnamefont{F.}~\bibnamefont{Setiawan}} \bibnamefont{and}
  \bibinfo{author}{\bibfnamefont{S.~D.} \bibnamefont{Sarma}},
  \bibinfo{journal}{arXiv preprint arXiv:1509.05067}  (\bibinfo{year}{2015}).

\bibitem[{\citenamefont{Weiss}(1999)}]{dhh-9}
\bibinfo{author}{\bibfnamefont{U.}~\bibnamefont{Weiss}},
  \emph{\bibinfo{title}{Quantum dissipative systems}},
  vol.~\bibinfo{volume}{10} (\bibinfo{publisher}{World Scientific},
  \bibinfo{year}{1999}).

\bibitem[{\citenamefont{Illes et~al.}(2015)\citenamefont{Illes, Roy, and
  Hughes}}]{dhh-10}
\bibinfo{author}{\bibfnamefont{E.}~\bibnamefont{Illes}},
  \bibinfo{author}{\bibfnamefont{C.}~\bibnamefont{Roy}}, \bibnamefont{and}
  \bibinfo{author}{\bibfnamefont{S.}~\bibnamefont{Hughes}},
  \bibinfo{journal}{Optica} \textbf{\bibinfo{volume}{2}}, \bibinfo{pages}{689}
  (\bibinfo{year}{2015}).

\bibitem[{\citenamefont{Silaev et~al.}(2014)\citenamefont{Silaev, Heikkil\"a,
  and Virtanen}}]{dhh-11}
\bibinfo{author}{\bibfnamefont{M.}~\bibnamefont{Silaev}},
  \bibinfo{author}{\bibfnamefont{T.~T.} \bibnamefont{Heikkil\"a}},
  \bibnamefont{and} \bibinfo{author}{\bibfnamefont{P.}~\bibnamefont{Virtanen}},
  \bibinfo{journal}{Phys. Rev. E} \textbf{\bibinfo{volume}{90}},
  \bibinfo{pages}{022103} (\bibinfo{year}{2014}).

\bibitem[{\citenamefont{Schaller}(2014)}]{dhh-12}
\bibinfo{author}{\bibfnamefont{G.}~\bibnamefont{Schaller}},
  \emph{\bibinfo{title}{Open Quantum Systems Far from Equilibrium}}
  (\bibinfo{publisher}{Springer, Lecture Notes in Physics},
  \bibinfo{year}{2014}).

\bibitem[{\citenamefont{Gumbs et~al.}(2015)\citenamefont{Gumbs, Iurov, and
  Horing}}]{ours}
\bibinfo{author}{\bibfnamefont{G.}~\bibnamefont{Gumbs}},
  \bibinfo{author}{\bibfnamefont{A.}~\bibnamefont{Iurov}}, \bibnamefont{and}
  \bibinfo{author}{\bibfnamefont{N.~J.~M.} \bibnamefont{Horing}},
  \bibinfo{journal}{Phys. Rev. B} \textbf{\bibinfo{volume}{91}},
  \bibinfo{pages}{235416} (\bibinfo{year}{2015}).

\bibitem[{\citenamefont{Singer et~al.}(2011)\citenamefont{Singer, Mecklenburg,
  White, and Regan}}]{dhh-14}
\bibinfo{author}{\bibfnamefont{S.~B.} \bibnamefont{Singer}},
  \bibinfo{author}{\bibfnamefont{M.}~\bibnamefont{Mecklenburg}},
  \bibinfo{author}{\bibfnamefont{E.~R.} \bibnamefont{White}}, \bibnamefont{and}
  \bibinfo{author}{\bibfnamefont{B.~C.} \bibnamefont{Regan}},
  \bibinfo{journal}{Phys. Rev. B} \textbf{\bibinfo{volume}{84}},
  \bibinfo{pages}{195468} (\bibinfo{year}{2011}).

\bibitem[{\citenamefont{Deshpande et~al.}(2009)\citenamefont{Deshpande, Hsieh,
  Bushmaker, Bockrath, and Cronin}}]{dhh-15}
\bibinfo{author}{\bibfnamefont{V.~V.} \bibnamefont{Deshpande}},
  \bibinfo{author}{\bibfnamefont{S.}~\bibnamefont{Hsieh}},
  \bibinfo{author}{\bibfnamefont{A.~W.} \bibnamefont{Bushmaker}},
  \bibinfo{author}{\bibfnamefont{M.}~\bibnamefont{Bockrath}}, \bibnamefont{and}
  \bibinfo{author}{\bibfnamefont{S.~B.} \bibnamefont{Cronin}},
  \bibinfo{journal}{Phys. Rev. Lett.} \textbf{\bibinfo{volume}{102}},
  \bibinfo{pages}{105501} (\bibinfo{year}{2009}).

\bibitem[{\citenamefont{Tessier et~al.}(2007)\citenamefont{Tessier, Bardoux,
  Boué, Filloy, and Fournier}}]{dhh-16}
\bibinfo{author}{\bibfnamefont{G.}~\bibnamefont{Tessier}},
  \bibinfo{author}{\bibfnamefont{M.}~\bibnamefont{Bardoux}},
  \bibinfo{author}{\bibfnamefont{C.}~\bibnamefont{Boué}},
  \bibinfo{author}{\bibfnamefont{C.}~\bibnamefont{Filloy}}, \bibnamefont{and}
  \bibinfo{author}{\bibfnamefont{D.}~\bibnamefont{Fournier}},
  \bibinfo{journal}{Applied Physics Letters} \textbf{\bibinfo{volume}{90}},
  \bibinfo{pages}{171112} (\bibinfo{year}{2007}).

\bibitem[{\citenamefont{Kucsko et~al.}(2013)\citenamefont{Kucsko, Maurer, Yao,
  Kubo, Noh, Lo, Park, and Lukin}}]{dhh-17}
\bibinfo{author}{\bibfnamefont{G.}~\bibnamefont{Kucsko}},
  \bibinfo{author}{\bibfnamefont{P.~C.} \bibnamefont{Maurer}},
  \bibinfo{author}{\bibfnamefont{N.~Y.} \bibnamefont{Yao}},
  \bibinfo{author}{\bibfnamefont{M.}~\bibnamefont{Kubo}},
  \bibinfo{author}{\bibfnamefont{H.~J.} \bibnamefont{Noh}},
  \bibinfo{author}{\bibfnamefont{P.~K.} \bibnamefont{Lo}},
  \bibinfo{author}{\bibfnamefont{H.}~\bibnamefont{Park}}, \bibnamefont{and}
  \bibinfo{author}{\bibfnamefont{M.~D.} \bibnamefont{Lukin}},
  \bibinfo{journal}{Nature} \textbf{\bibinfo{volume}{500}},
  \bibinfo{pages}{7460} (\bibinfo{year}{2013}).

\bibitem[{\citenamefont{Mecklenburg et~al.}(2015)\citenamefont{Mecklenburg,
  Hubbard, White, Dhall, Cronin, Aloni, and Regan}}]{dhh-18}
\bibinfo{author}{\bibfnamefont{M.}~\bibnamefont{Mecklenburg}},
  \bibinfo{author}{\bibfnamefont{W.~A.} \bibnamefont{Hubbard}},
  \bibinfo{author}{\bibfnamefont{E.~R.} \bibnamefont{White}},
  \bibinfo{author}{\bibfnamefont{R.}~\bibnamefont{Dhall}},
  \bibinfo{author}{\bibfnamefont{S.~B.} \bibnamefont{Cronin}},
  \bibinfo{author}{\bibfnamefont{S.}~\bibnamefont{Aloni}}, \bibnamefont{and}
  \bibinfo{author}{\bibfnamefont{B.~C.} \bibnamefont{Regan}},
  \bibinfo{journal}{Science} \textbf{\bibinfo{volume}{347}},
  \bibinfo{pages}{629} (\bibinfo{year}{2015}).

\bibitem[{\citenamefont{Kibis}(2010)}]{Kibis}
\bibinfo{author}{\bibfnamefont{O.~V.} \bibnamefont{Kibis}},
  \bibinfo{journal}{Phys. Rev. B} \textbf{\bibinfo{volume}{81}},
  \bibinfo{pages}{165433} (\bibinfo{year}{2010}).

\bibitem[{\citenamefont{Busl et~al.}(2012)\citenamefont{Busl, Platero, and
  Jauho}}]{Busl}
\bibinfo{author}{\bibfnamefont{M.}~\bibnamefont{Busl}},
  \bibinfo{author}{\bibfnamefont{G.}~\bibnamefont{Platero}}, \bibnamefont{and}
  \bibinfo{author}{\bibfnamefont{A.-P.} \bibnamefont{Jauho}},
  \bibinfo{journal}{Physical Review B} \textbf{\bibinfo{volume}{85}},
  \bibinfo{pages}{155449} (\bibinfo{year}{2012}).

\bibitem[{\citenamefont{Giovannetti et~al.}(2008)\citenamefont{Giovannetti,
  Khomyakov, Brocks, Karpan, van~den Brink, and Kelly}}]{PR1}
\bibinfo{author}{\bibfnamefont{G.}~\bibnamefont{Giovannetti}},
  \bibinfo{author}{\bibfnamefont{P.~A.} \bibnamefont{Khomyakov}},
  \bibinfo{author}{\bibfnamefont{G.}~\bibnamefont{Brocks}},
  \bibinfo{author}{\bibfnamefont{V.~M.} \bibnamefont{Karpan}},
  \bibinfo{author}{\bibfnamefont{J.}~\bibnamefont{van~den Brink}},
  \bibnamefont{and} \bibinfo{author}{\bibfnamefont{P.~J.} \bibnamefont{Kelly}},
  \bibinfo{journal}{Phys. Rev. Lett.} \textbf{\bibinfo{volume}{101}},
  \bibinfo{pages}{026803} (\bibinfo{year}{2008}).

\bibitem[{\citenamefont{Grosse et~al.}(2011)\citenamefont{Grosse, Bae, Lian,
  Pop, and King}}]{PR2}
\bibinfo{author}{\bibfnamefont{K.~L.} \bibnamefont{Grosse}},
  \bibinfo{author}{\bibfnamefont{M.-H.} \bibnamefont{Bae}},
  \bibinfo{author}{\bibfnamefont{F.}~\bibnamefont{Lian}},
  \bibinfo{author}{\bibfnamefont{E.}~\bibnamefont{Pop}}, \bibnamefont{and}
  \bibinfo{author}{\bibfnamefont{W.~P.} \bibnamefont{King}},
  \bibinfo{journal}{Nature nanotechnology} \textbf{\bibinfo{volume}{6}},
  \bibinfo{pages}{287} (\bibinfo{year}{2011}).

\bibitem[{\citenamefont{Reckinger et~al.}(2013)\citenamefont{Reckinger, Vlad,
  Melinte, Colomer, and Sarrazin}}]{PR3}
\bibinfo{author}{\bibfnamefont{N.}~\bibnamefont{Reckinger}},
  \bibinfo{author}{\bibfnamefont{A.}~\bibnamefont{Vlad}},
  \bibinfo{author}{\bibfnamefont{S.}~\bibnamefont{Melinte}},
  \bibinfo{author}{\bibfnamefont{J.-F.} \bibnamefont{Colomer}},
  \bibnamefont{and} \bibinfo{author}{\bibfnamefont{M.}~\bibnamefont{Sarrazin}},
  \bibinfo{journal}{Applied Physics Letters} \textbf{\bibinfo{volume}{102}},
  \bibinfo{pages}{211108} (\bibinfo{year}{2013}).

\bibitem[{\citenamefont{Shung}(1986)}]{W5}
\bibinfo{author}{\bibfnamefont{K.~W.~K.} \bibnamefont{Shung}},
  \bibinfo{journal}{Phys. Rev. B} \textbf{\bibinfo{volume}{34}},
  \bibinfo{pages}{979} (\bibinfo{year}{1986}).

\bibitem[{\citenamefont{Ando}(2006)}]{Ando}
\bibinfo{author}{\bibfnamefont{T.}~\bibnamefont{Ando}},
  \bibinfo{journal}{Journal of the Physical Society of Japan}
  \textbf{\bibinfo{volume}{75}}, \bibinfo{pages}{074716}
  (\bibinfo{year}{2006}).

\bibitem[{\citenamefont{Maldague}(1978)}]{15}
\bibinfo{author}{\bibfnamefont{P.~F.} \bibnamefont{Maldague}},
  \bibinfo{journal}{Surface Science} \textbf{\bibinfo{volume}{73}},
  \bibinfo{pages}{296 } (\bibinfo{year}{1978}).

\bibitem[{\citenamefont{Baorong et~al.}(2009)\citenamefont{Baorong, Linghua,
  Jianhong, and Xiwei}}]{Yan}
\bibinfo{author}{\bibfnamefont{Y.}~\bibnamefont{Baorong}},
  \bibinfo{author}{\bibfnamefont{K.}~\bibnamefont{Linghua}},
  \bibinfo{author}{\bibfnamefont{L.}~\bibnamefont{Jianhong}}, \bibnamefont{and}
  \bibinfo{author}{\bibfnamefont{H.}~\bibnamefont{Xiwei}},
  \bibinfo{journal}{Plasma Science and Technology}
  \textbf{\bibinfo{volume}{11}}, \bibinfo{pages}{515} (\bibinfo{year}{2009}).

\bibitem[{\citenamefont{Horing}(2009)}]{Horing2009}
\bibinfo{author}{\bibfnamefont{N.~J.~M.} \bibnamefont{Horing}},
  \bibinfo{journal}{Physical Review B} \textbf{\bibinfo{volume}{80}},
  \bibinfo{pages}{193401} (\bibinfo{year}{2009}).

\end{thebibliography}
\end{document}